\documentclass[sigplan,nonacm]{acmart}

\usepackage{algorithm,algpseudocode,amsmath,booktabs,makecell}
\usepackage{graphicx}
\usepackage{subcaption}
\usepackage{microtype}
\usepackage{lipsum}
\usepackage{soul}
\usepackage[nohyperlinks]{acronym}
\usepackage{enumitem} % for leftmargin control on itemize

\AtBeginDocument{%
  }

\newcommand\blfootnote[1]{%
  \begingroup
  \renewcommand\thefootnote{}\footnote{#1}%
  \addtocounter{footnote}{-1}%
  \endgroup
}

\newacro{aimc}[AIMC]{Analog In-Memory Computing}
\newacro{3dimc}[3D-IMC]{3D AIMC}
\newacro{camp}[CAMP]{Context-Aware Mapping and Placement}
\newacro{cnn}[CNN]{Convolutional Neural Network}
\newacro{dnn}[DNN]{Deep Neural Network}
\newacro{dpu}[DPU]{Digital Processing Unit}
\newacro{dram}[DRAM]{Dynamic Random-Access Memory}
\newacro{ep}[EP]{Expert Parallelism}
\newacro{ffn}[FFN]{Feed-Forward Network}
\newacro{gpu}[GPU]{Graphics Processing Unit}
\newacro{hbm}[HBM]{High-Bandwidth Memory}
\newacro{ir}[IR]{Intermediate Representation}
\newacro{kv}[KV]{Key-Value}
\newacro{lg}[LG]{Leaf Group}
\newacro{llm}[LLM]{Large Language Model}
\newacro{mha}[MHA]{Multi-Head Attention}
\newacro{milp}[MILP]{Mixed-Integer Linear Programming}
\newacro{moe}[MoE]{Mixture-of-Experts}
\newacro{mvm}[MVM]{Matrix-Vector Multiplication}
\newacro{nvm}[NVM]{Non-Volatile Memory}
\newacro{ott}[OTT]{One-Tier-at-a-Time}
\newacro{pcm}[PCM]{Phase-Change Memory}
\newacro{pci}[PCIe]{Peripheral Component Interconnect express}
\newacro{pim}[PIM]{Processing-In-Memory}
\newacro{bp}[BP]{Batch Parallelism}
\newacro{qa}[QA]{Question Answering}
\newacro{qkv}[QKV]{Query-Key-Value}
\newacro{g+u}[G+U]{Gated Up}
\newacro{rram}[RRAM]{Resistive Random-Access Memory}
\newacro{slo}[SLO]{Service-Level Objective}
\newacro{ssm}[SSM]{State-Space Model}
\newacro{tp}[TP]{Tensor Parallelism}
\newacro{tpu}[TPU]{Tensor Processing Unit}
\newacroplural{lg}[LGs]{Leaf Groups}
\newacroplural{llm}[LLMs]{Large Language Models}
\newacroplural{moe}[MoEs]{Mixtures-of-Experts}
\newacroplural{mvm}[MVMs]{Matrix-Vector Multiplications}
\newacroplural{ssm}[SSMs]{State-Space Models}

\begin{document}

\title{CAMP: \texorpdfstring{\underline{C}}{C}ontext-\texorpdfstring{\underline{A}}{A}ware \texorpdfstring{\underline{M}}{M}apping and \texorpdfstring{\underline{P}}{P}lacement for LLMs on 3D-IMC Arrays}

\author{Orhun Görkem}
\affiliation{\institution{IBM Research}
  \city{R\"uschlikon}
  \country{Switzerland}}
\email{orhun.goerkem1@ibm.com}

\author{Athanasios Vasilopoulos}
\affiliation{\institution{IBM Research}
  \city{R\"uschlikon}
  \country{Switzerland}}
\email{ATV@zurich.ibm.com}

\author{Georgios Pagonis}
\affiliation{\institution{IBM Research}
  \city{R\"uschlikon}
  \country{Switzerland}}
\email{Georgios.Pagonis@ibm.com}

\author{Victoria Clerico}
\affiliation{\institution{IBM Research}
  \city{R\"uschlikon}
  \country{Switzerland}}
\email{Victoria.Clerico.Da.Costa@ibm.com}

\author{Hadjer Benmeziane}
\affiliation{\institution{IBM Research}
  \city{R\"uschlikon}
  \country{Switzerland}}
\email{Hadjer.Benmeziane@ibm.com}

\author{Irem Boybat}
\affiliation{\institution{IBM Research}
  \city{R\"uschlikon}
  \country{Switzerland}}
\email{IBO@zurich.ibm.com}

\author{Hsinyu Tsai}
\affiliation{\institution{IBM Research}
  \city{San Jose}
  \state{CA}
  \country{USA}}
\email{htsai@us.ibm.com}

\author{Abu Sebastian}
\affiliation{\institution{IBM Research}
  \city{R\"uschlikon}
  \country{Switzerland}}
\email{ASE@zurich.ibm.com}

\author{William Andrew Simon}
\affiliation{\institution{IBM Research}
  \city{R\"uschlikon}
  \country{Switzerland}}
\email{William.Simon1@ibm.com}

\begin{abstract}
\Acp{llm} incur severe memory-access overheads due to the cost of loading large weight matrices and \ac{kv} caches.
\Ac{aimc} is an emerging solution that enables computation directly on static weights stored in memory, substantially reducing energy and latency.
\acs{3dimc} further stacks memory arrays into vertical tiers, increasing capacity but revealing a new challenge: \textit{how} weights are placed in memory determines which \acp{mvm} can execute concurrently, affecting latency and throughput.
The concurrency aspect of \acs{3dimc} weight-mapping remains underexplored in the literature, leading to sub-optimal utilization of compute resources on \acs{3dimc} chips.
%However, \textit{how} weights are mapped in memory significantly affects latency and throughput, presenting an optimization opportunity underexplored in the literature.

In this context we propose \ac{camp}, an inference-context-aware weight mapping algorithm for \acs{3dimc}-based \ac{llm} accelerators. \Ac{camp} captures the potential resource contention resulting from the architecture, targeted network, and runtime context in a \textit{conflict matrix}, enabling throughput-optimized weight mapping.
This mapping is refined through a profiling-driven feedback loop that efficiently prunes the search space and returns optimized mappings within practical runtimes.
\Ac{camp} is an offline, software-based solution requiring no hardware changes and improving \acs{3dimc} compute utilization from 42.4\% to 60.3\%, delivering up to $2.0\times$ higher \acs{3dimc} throughput and $1.74\times$ higher end-to-end throughput over state-of-the-art mapping algorithms.

\end{abstract}
\begin{CCSXML}
<ccs2012>
   <concept>
       <concept_id>10010520.10010521.10010542.10010546</concept_id>
       <concept_desc>Computer systems organization~Heterogeneous (hybrid) systems</concept_desc>
       <concept_significance>500</concept_significance>
       </concept>
   <concept>
       <concept_id>10010583.10010786.10010787.10010788</concept_id>
       <concept_desc>Hardware~Emerging architectures</concept_desc>
       <concept_significance>500</concept_significance>
       </concept>
   <concept>
   %     <concept_id>10010583.10010786.10010809</concept_id>
   %     <concept_desc>Hardware~Memory and dense storage</concept_desc>
   %     <concept_significance>300</concept_significance>
   %     </concept>
       <concept_id>10010520.10010521.10010528</concept_id>
       <concept_desc>Computer systems organization~Parallel architectures</concept_desc>
       <concept_significance>500</concept_significance>
       </concept>
 </ccs2012>
\end{CCSXML}

\ccsdesc[500]{Computer systems organization~Heterogeneous (hybrid) systems}
\ccsdesc[500]{Hardware~Emerging architectures}
%\ccsdesc[300]{Hardware~Memory and dense storage}
\ccsdesc[500]{Computer systems organization~Parallel architectures}

\keywords{In-memory computing; LLM inference; batch-parallelism; weight mapping; 3D-IMC}


\maketitle

% Expand acronyms again on first use in the paper body because the abstract
% must remain self-contained.
\acresetall

\section{Introduction}

\blfootnote{An open-source script for simulations can be found at \\ \nolinkurl{https://anonymous.4open.science/r/3DIMC_CM_Mapping-904D/.}}

\Acp{llm} have recently demonstrated remarkable capabilities across natural language processing and beyond~\cite{naveed2024}, making them widely adopted in industry and everyday life.
Serving \acp{llm} efficiently is therefore an important but challenging goal, made harder by growing parameter counts~\cite{naveed2024,zoph2022,li2025}.

The mainstream hardware for \ac{llm} inference is the \ac{gpu}.
While well suited to the compute-heavy prefill phase, \ac{gpu} compute resources are underutilized during the memory-bound decode phase~\cite{li2025,kim2025,heo2024}, where limited memory bandwidth results in \ac{kv} cache and weight loading dominating inference time.
Prior works target the \ac{kv}-loading bottleneck via software techniques such as \ac{kv} compression~\cite{zandieh2026,xiao2024,sun2025}, new models such as \acp{ssm}~\cite{gu2022}, or memory-centric hardware such as \ac{dram}-based \ac{pim}~\cite{xu2026,seo2025,gu2025,kim2025}.
Meanwhile, the weight loading bottleneck can be mitigated with sparsity and quantization~\cite{guo2024,mozaffari2025}, or eliminated altogether via \ac{nvm}-based \ac{aimc} accelerators~\cite{bavandpour2019,Buchel2025,bommana2025,Meng2025,dhingra2025,tsai2025}.

\begin{figure*}[t]
    \centering
    \includegraphics[width = 2.0\columnwidth]{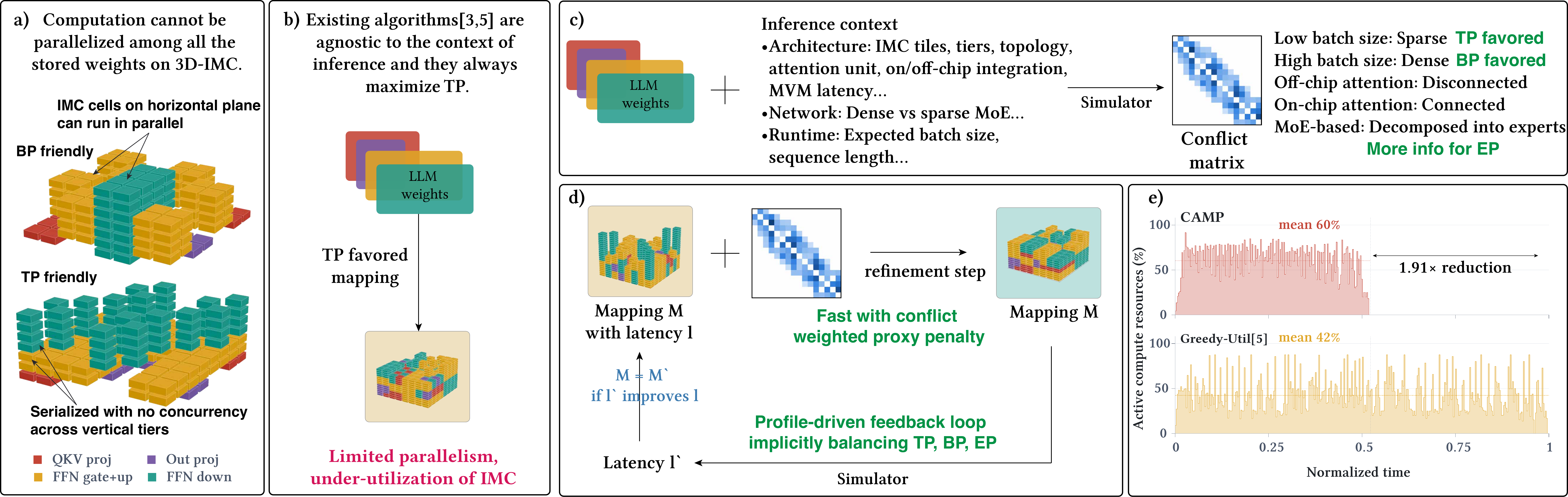}
    \Description{Five-panel overview.
    Panel (a) contrasts a batch-parallel mapping, which places differently colored \acs*{qkv}, output, \acs*{g+u}, and down-projection weights on disjoint horizontal resources, with a tensor-parallel mapping that spreads weights horizontally but serializes work across vertical tiers.
    Panel (b) shows that a context-agnostic, tensor-parallel mapping underutilizes the \acs*{3dimc} macro.
    Panels (c) and (d) show \acs*{camp} converting architecture, network, and runtime context into a conflict matrix, then repeatedly refining and simulating a mapping.
    Panel (e) compares active compute resources over normalized time: \acs*{camp} averages 60 percent utilization versus 42 percent for Greedy-Util and completes the work in roughly half the time.}
    \caption{ (a) Static weights may be mapped to improve the latency of an individual \acs*{mvm} (\acs*{tp}) or enable \acsp*{mvm} to execute concurrently (\acs*{bp}).
    (b) Existing algorithms always maximize \acs*{tp} agnostically to inference context.
    (c) In contrast, \acs*{camp} maps in a context-aware manner by first encoding the context in a conflict matrix.
    (d) A refinement loop converges to a mapping with an optimized balance of \acs*{tp}, \acs*{bp}, and \acs*{ep}, (e) improving runtime efficiency via better utilization of \acs*{3dimc} compute resources.}
    \label{fig:intro}
\end{figure*}

% Complementarily, \ac{nvm}-based \ac{imc} accelerators eliminate \ac{mvm} weight loading~\cite{bavandpour2019,Buchel2025,bommana2025,Meng2025,dhingra2025,tsai2025}.

\Ac{aimc} accelerators eliminate weight movement by executing \acp{mvm} involving static weights directly within the memory array.
However, \ac{nvm}'s poor write and endurance characteristics~\cite{jain2023} require all network weights to fit on-chip.
% \Ac{imc} accelerators perform \acp{mvm} directly within the memory arrays that store static weights, eliminating weight loading but requiring all network weights to fit on-chip.
As model size grows, this constraint has encouraged research into 3D \ac{aimc} devices, whose memory arrays consist of vertically stacked tiers, enabling high storage density~\cite{Buchel2025,tsai2025,dhingra2025,Meng2025}.

% Crucially though, increasing vertical capacity \textbf{does not} increase compute capability, which is limited to the total array count irrespective of vertical integration.
While \ac{3dimc} increases storage capacity via vertical integration, compute capability remains static, as multiple tiers within the same array cannot be accessed simultaneously.
This limitation creates a trade-off between prioritizing single-\ac{mvm} latency and enabling inter-\ac{mvm} concurrency, which can be tuned through the weight mapping as illustrated in Figure~\ref{fig:intro}a.
% However, compute density does not scale at all with the storage density in such devices~\cite{Buchel2025} because only one vertical tier of memory can be activated for read/write/compute operations concurrently.
% This limitation necessitates algorithms utilize the available \ac{3dimc} compute resources efficiently.
% A primary determinant of resource utilization on \ac{3dimc} accelerators is therefore how weight matrices are mapped onto the memory array~\cite{Meng2025,houshmand2024,Buchel2025,bavandpour2019,huang2025}.
%, as \textbf{weight mapping} jointly controls various parallelism opportunities in the inference.
Weight mapping determines the degree of \textbf{\ac{tp}} by fixing how many compute resources are allocated to a single \ac{mvm}, tuning its latency.
It also determines the degree of \textbf{\ac{bp}}, the ability of the accelerator to process multiple tokens within a batch concurrently by assigning \acp{mvm} to disjoint resources and enabling inter-token computation overlap.
Finally, for \ac{moe} models, weight mapping defines \textbf{\ac{ep}}, or how compute resources are allocated to experts within an \ac{moe} layer.

Although optimized weight mapping benefits runtime performance due to better compute utilization, optimizing \ac{3dimc} weight mapping is challenging:  

\noindent\textbf{Optimizing individual parallelism dimensions} by finding a mapping that maximizes even a single form of parallelism is difficult due to varying \ac{llm} workload structures, runtime contexts, and accelerator architectures.

\noindent\textbf{Balancing parallelism} across the \ac{tp}, \ac{bp}, and \ac{ep} paradigms is another challenge, as optimizing one form of parallelism can directly constrain the others~\cite{hidayetoglu2026}.
For example, increasing \ac{tp} allocates more resources to a single \ac{mvm}, reducing available resources for concurrently executing other \acp{mvm} and thereby limiting \ac{bp}.
Similarly, increasing \ac{ep} concentrates resources in experts, allowing experts to run concurrently but reducing availability to other \acp{mvm}.

\noindent\textbf{Robustness across inference contexts} consisting of system architectures, varying workloads, and expected runtime contexts is necessary to develop a generally applicable performant mapping solution.  

\noindent Finally, \textbf{mapping-space complexity} precludes exhaustive search due to the enormous weight mapping search space, which grows rapidly with the number of \acp{mvm} to be mapped and the number of available resource configurations along both the planar and vertical dimensions.
State-of-the-art mapping methods~\cite{Buchel2025,bavandpour2019} overcome this by greedy mapping procedures only maximizing \ac{tp}, at the expense of \ac{bp}, \ac{ep}, and robustness against varying inference contexts.

% only maximizing \ac{tp} at the expense of \ac{bp} and \ac{ep}, are not robust across inference contexts, and use greedy mapping approaches to overcoming mapping-space complexity at the cost of mapping performance.
% State-of-the-art mapping methods~\cite{Buchel2025,bavandpour2019} use greedy mapping approaches with linear computational complexity, but remain far from throughput-optimal and lack robustness across workloads.

To address these challenges, we propose \textbf{\ac{camp}}, an inference-context-aware mapping algorithm for \ac{3dimc}-based \ac{llm} accelerators.
Figure~\ref{fig:intro} provides an overview of \ac{camp}.
% First, a profiling phase analyzes the workload under the expected inference context, consisting of the system architecture, target \ac{llm}, and runtime context, and encodes the probability of resource conflicts between pairs of \acp{mvm} into a \textbf{conflict matrix}.
% The subsequent phase maps weights to the \ac{3dimc} macro, and refines the mapping via a \textbf{profile-guided, feedback-oriented pruned search} over possible weight reshapings and placements to identify the throughput-maximizing mapping configuration.
\Ac{camp} addresses the challenges of weight mapping as follows:

(1) First, a profiling phase analyzes the workload under the expected inference context — the system architecture, target \ac{llm}, and runtime context — and encodes the probability of resource conflicts between pairs of static \acp{mvm} into a \textbf{conflict matrix}.
The mapping process then uses this matrix to adapt automatically and robustly to the given circumstances.
% Because the mapping process is built on top of this matrix, it adjusts automatically to the given circumstances, making it adaptive and robust.
    % \item The conflict matrix mathematical abstraction encodes the context in which inference is run: the system architecture and constraints, model details, and expected runtime context such as batch size and sequence length.
    % Because the mapping process is built on top of this matrix, it adjusts automatically to the given circumstances, making it adaptive and robust.

(2) Next, an initial mapping phase maps all static \acp{mvm} to the \ac{3dimc} accelerator, using the conflict matrix to map \acp{mvm} likely to contend for resources to disjoint resource groups, improving \ac{bp} and \ac{ep}.

% \noindent(2) Next, an initial mapping phase utilizes the conflict matrix to identify pairs of \acp{mvm} likely to contend for the same resources.
% Whenever possible, such pairs are mapped to disjoint resource groups, improving \ac{bp} and \ac{ep}.

(3) Finally, the mapping is refined via \textbf{a profile-guided, feedback-driven pruned search}, which utilizes the projected latency of varied candidate mappings to guide refinement toward operating points at which the balance between \ac{tp}, \ac{bp}, and \ac{ep} is tuned for maximum resource utilization.

The exact planar and vertical weight placements are obtained efficiently through a proxy penalty computed from the conflict matrix, which prunes a large region of the search space.
A beam search strategy trades solution optimality against optimization time, placing \ac{camp} between a fully greedy heuristic and globally optimal methods, while yielding near-optimal mappings within acceptable runtimes.

The contributions of this paper are thus:
\begin{itemize}[leftmargin=*]
    \item We analyze how system architecture, \ac{llm} model, and runtime characteristics, collectively referred to as the \textbf{inference context}, impact runtime throughput and latency.
    \item We introduce \textbf{the conflict matrix}, a compact abstract representation that encodes the inference context in a single structure amenable to efficient optimization.
    \item We propose \textbf{\Acf{camp}}, an inference-context-aware weight-mapping algorithm built on this abstraction, which first determines each weight tensor's placement shape and assignment to \ac{3dimc}, then refines placement via a pruned search over reshapings, guided by profiling-in-the-loop feedback.
    \item We demonstrate \textbf{cross-regime adaptivity}, as \ac{camp} produces markedly different mappings across runtime contexts, networks, and hardware, minimizing latency at low batch sizes, maximizing throughput at high batch sizes, and natively enabling \ac{ep} in \ac{moe} \acp{llm}.
\end{itemize}
Thorough evaluations demonstrate up to $2.0\times$ higher \ac{3dimc} runtime throughput, $1.74\times$ higher end-to-end throughput, and a 42\% increase in \ac{3dimc} compute utilization, validating \ac{camp}'s \ac{3dimc} weight mapping effectiveness.
% An optimality analysis demonstrates that our approach closes $xxx\%$ of the gap between SotA mapping algorithms and the theoretical throughput upper bound of the target \ac{3dimc} accelerator.

% \clearpage

\section{Background}

\subsection{Large Language Models}

\Acp{llm} are state-of-the-art models across many domains and comprise \textit{transformer blocks}, which employ the \textit{attention} mechanism, capturing long-range inter-token dependencies~\cite{naveed2024}, and \ac{ffn} layers that transform the learned token-level associations into higher-level representations.
While \acp{llm}' substantial representational capacity makes them extremely versatile, they also incur substantial compute and memory demands from their static and dynamic parameter counts.

\subsection{Challenges in LLM execution}

The effectiveness of the attention mechanism on long sequences stems from its ability to model dependencies between all token pairs~\cite{vaswani2017}.
However, the quadratic computational complexity of attention poses a major challenge for long-context inference.
\Ac{kv} caching mitigates this issue by computing and caching keys and values for all tokens once (\textit{prefill}) and reusing the stored values for each output token (\textit{decode}), reducing per-token attention complexity from $O(N^2)$ to $O(N)$~\cite{shazeer2019}.
% inference is split into a prefill phase, in which the keys and values for all input tokens are computed once and stored, and a decode phase, in which these cached entries are reused for each output token, reducing per-token attention complexity from $O(N^2)$ to $O(N)$.

While \ac{kv} caching reduces decode-phase computation, cached values must be re-fetched from the memory for every generated token.
Large weight parameter counts add further memory pressure, motivating memory-centric solutions to mitigate bandwidth limitations.
% Indeed, vLLM partitions \ac{gpu} cache in an \hl{N}:\hl{M} ratio between storage for \ac{kv} cache and static weights, respectively.
% The \ac{kv} cache and static weights also have different characteristics: the \ac{kv} cache is dynamically generated and unique to each token, whereas weights are fixed after training and shared across tokens.
% These fundamental differences motivate distinct solutions for each regime; for example, \ac{kv}-cache compression~\cite{zandieh2026,xiao2024} and \acp{ssm}~\cite{gu2022} alleviate \ac{kv} memory pressure, 
% whereas \ac{moe} models increase MVM parameter count without a commensurate increase in bandwidth or compute requirements~\cite{shazeer2017}.
% These complementary optimizations motivate a more fundamental shift toward \textit{attention/\ac{mvm} disaggregation}, in which attention and \acp{mvm} execute on separate specialized hardware.
% Disaggregation has been studied across a range of architectures and accelerators~\cite{aubrey2026,park2024,pan2025,xu2026,yu2024}.

% Although NVIDIA's new \ac{gpu}-GROQ architecture is a prominent example~\cite{aubrey2026}, disaggregation has been studied across a range of architectures and accelerators~\cite{aubrey2026,park2024,pan2025,xu2026,yu2024,duan2026}.

\subsection{Reducing data movement via IMC}
\label{sec:imc}

% In- and near-memory computing paradigms have emerged as promising approaches to model disaggregation and to reduce the memory pressure imposed by large \ac{kv} caches and static weights.
In- and near-memory computing paradigms have emerged as promising approaches to reduce the memory pressure imposed by large \ac{kv} caches and static weights.
Their key principle is to minimize expensive data transfers by performing computation closer to where data are stored or by exploiting high-bandwidth local interconnects.

On the attention side, one such approach is \ac{dram}-based \ac{pim}~\cite{park2024,huang2025,seo2025,qu2025,pan2025,xu2026}, alleviating the \ac{kv}-cache access bottleneck by exploiting the internal bandwidth of \ac{dram} and performing attention-related computation near \ac{kv} storage.

\begin{figure}[t]
    \centering
    \includegraphics[width = 0.85\columnwidth]{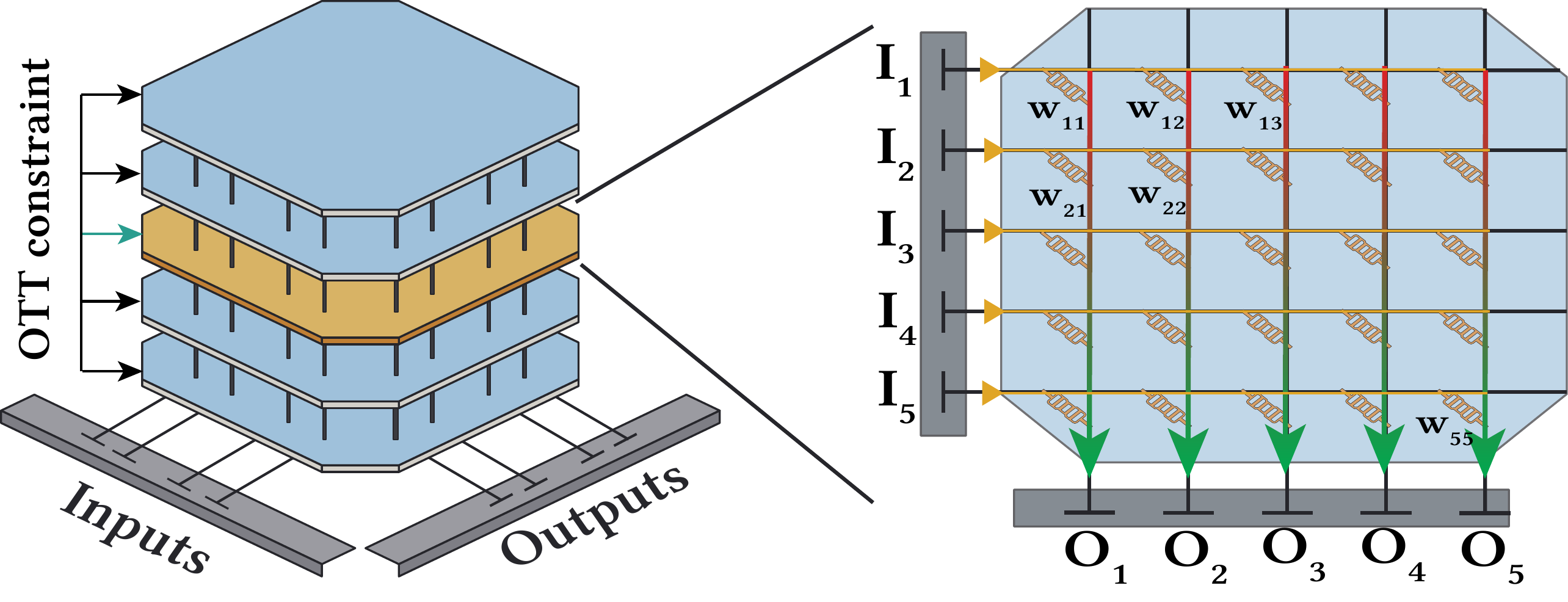}
    \Description{Figure illustrating a \acs*{3dimc} stack as five tiers of memory arrays.
    The middle array is blown up to highlight the internal crossbar which contains weight values.
    Arrows indicate how input activations flow into the array, and output activations flow out.}
    \caption{Each tier in an \acs*{3dimc} memory array stores weights that are used to perform \acsp*{mvm} in-situ.}
    \label{fig:aimc}
\end{figure}

A complementary approach addressing static weights is \ac{nvm}-based \ac{aimc} accelerators, as illustrated in Figure~\ref{fig:aimc}.
Such architectures represent model weights as analog states such as cell conductance and exploit analog device characteristics to perform massively parallel \acp{mvm} directly within memory arrays, eliminating weight movement~\cite{jain2023,sebastian2020,verma2019,Buchel2025,bommana2025,shao2019,Meng2025,dhingra2025,shim2021,jang2025}.
While \ac{aimc}'s poor write and endurance characteristics~\cite{jain2023} make it ill-suited to storing frequently changing data such as the \ac{kv} cache, avoiding repeated static weight transfers allows \ac{aimc} accelerators to reduce inference latency and improve energy efficiency significantly.
The different characteristics of \ac{kv} cache and static weights have motivated interest in \textit{attention/static \ac{mvm} disaggregation}, in which attention operations and \acp{mvm} with static weights execute on separate specialized hardware~\cite{aubrey2026,park2024,pan2025,xu2026,yu2024}.

For conciseness, we hereafter simply use the term \ac{mvm} to refer to all \acp{mvm} with static weights accelerated with \ac{aimc}, specifically the \ac{qkv}, output, and the \ac{ffn} \ac{g+u} and down projections.
This is in contrast to \acp{mvm} with dynamic inputs occurring in the attention block and executed on the attention unit.
% We use \textit{\acp{mvm}} to refer to all weight-stationary \acp{mvm} accelerated with \ac{aimc}, including the \ac{qkv}, output, and the \ac{ffn} \ac{g+u} and down projections.
% This term is broader than \ac{ffn}, which refers only to the feed-forward sublayer of a transformer block.
% AIMC accelerators therefore focus on the \ac{mvm} operations of LLM inference.

% Because AIMC computes on weights in place, model weights must be mapped onto and programmed into IMC memory arrays prior to inference.
% Efficient mapping and placement of LLM weights onto IMC arrays have therefore become important research challenges, encompassing issues such as array utilization, quantization, and data placement~\cite{bommana2025,bavandpour2019,shao2019,Kao2020,Meng2025,dhingra2025,houshmand2024,Buchel2025,tsai2025}.

\begin{figure*}[htbp]
    \centering
    \includegraphics[width=0.95\textwidth]{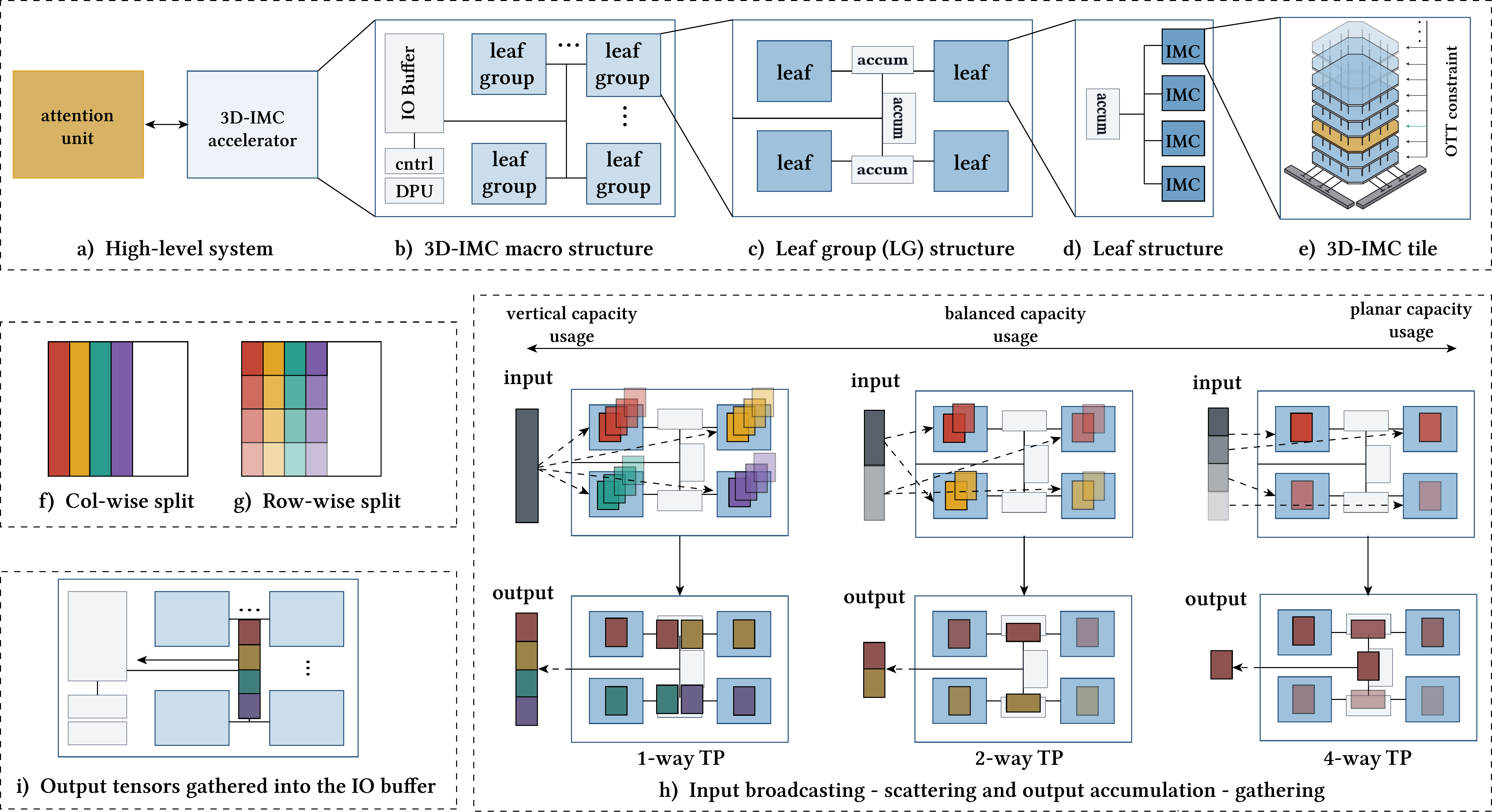}
    \Description{Nine-panel architecture and dataflow diagram.
    Panels (a)--(e) zoom from an attention unit connected to an \acs*{3dimc} macro, through the macro's H-tree of leaf groups, to leaves, accumulators, and a vertically stacked \acs*{3dimc} tile subject to the one-tier-at-a-time constraint.
    Panels (f) and (g) split a weight matrix by output columns and then input rows.
    Panel (h) compares one-, two-, and four-way tensor-parallel placements: increasing the degree shifts capacity use from vertical tiers to the planar dimension, broadcasts input chunks to more leaves, and accumulates more partial outputs.
    Panel (i) gathers the resulting output tensors into the global IO buffer.}
    \caption{Hierarchical H-tree architecture (a-e)~\cite{shim2021,shim2021_2,jang2025}.
             Weight mapping options when leaf groups comprise 4 leaves (f-i).}
    \label{fig:architecture_dataflow}
\end{figure*}

\subsection{3D-IMC opportunities and challenges}
\label{sec:3dimc}

% \begin{figure}[t]
%     \centering
%     \includegraphics[width = 0.98\columnwidth,trim={0.5cm 1cm 1cm 3.5cm},clip]{figures/active_weights_motivation.pdf}
%     \caption{
%         Potential active-weight capacity on 3D-IMC (dashed orange line) compared with the actual active weights engaged during Llama 3.1-70B inference under the mapping algorithm from~\cite{Buchel2025} (green line).
%         At a saturating batch size of 8, MVM computation uses only $\approx 42\%$ of the full 3D-IMC compute capacity on average.
%     }
%     \label{fig:active_weights}
% \end{figure}

As \acp{llm} continue to grow in parameter count, increasing memory density has become necessary to accommodate their storage requirements.
To address this capacity bottleneck, proposed three-dimensional \ac{aimc} (\ac{3dimc}) architectures significantly increase the weight capacity of conventional \ac{aimc} architectures through the vertical integration of memory arrays~\cite{Buchel2025,tsai2025,shim2021_2,dhingra2026,Meng2025}.

However, because vertical tier-level parallelism cannot be implemented due to peripheral sharing and vertical interconnect bandwidth constraints, compute capacity scales only in the planar dimension in several popular 3D-memory paradigms such as NAND flash, a limitation referred to as the \emph{\ac{ott}} constraint~\cite{Buchel2025,bavandpour2019}.

This limitation requires thoughtful interconnect design to efficiently serve activations and gather outputs.
A popular design that achieves these goals is a hierarchical H-tree~\cite{jang2025, dhingra2026, xu2026,pan2025}, illustrated in Figure~\ref{fig:architecture_dataflow}(a-e).
In such a design, the \ac{3dimc} architecture comprises an array of $N_{LG}$ \textit{\acp{lg}}, hereafter referred to as the \ac{3dimc} macro, interconnected by an H-tree topology~\cite{jang2025,shim2021,shim2021_2}.
This structure is preferred for its input-broadcasting and output-accumulation capabilities, enabling large \acp{mvm} to be parallelized across multiple \acp{lg}.
Each \ac{lg} is further composed of $N_{L}$ \textit{leaves}, each containing $N_{T}$ vertically stacked \ac{aimc} \textit{tiles}, as shown in Figure~\ref{fig:architecture_dataflow}(e).
Accumulation units within leaves and \acp{lg} facilitate output accumulation, enabling multiple leaves to contribute to the output of one \ac{mvm}, a feature exploited in Section~\ref{subsec:tiling}.
% \Acp{lg} also contain a \ac{dpu} and control unit for auxiliary operations such as activation functions.
Beyond the macro, the architecture provides a global IO buffer for intermediate vectors, a \ac{dpu} for activation functions, and a global controller.
The accelerator is paired with an attention unit which performs the attention stage, including \ac{kv}-cache loading, while the accelerator is responsible for the static \acp{mvm}.

% For brevity of results, we use $N_{LG} = 64$, $N_{L} = 4$, and $N_{T} = 4$; Appendix~\ref{app:arch_topology} evaluates other topologies.

% Nevertheless, 3D-IMC remains a promising solution for accelerating LLM inference due to the improvements in weight capacity and memory bandwidth it provides.
% Figure~\ref{fig:active_weights} suggests that the active weight compute capacity of standard 3D-IMC architectures is not far behind the size of the FFN layers of Llama 70B.
% However, existing mapping algorithms~\cite{Buchel2025,bavandpour2019} underutilize these resources.
% The imbalance between storage and compute capacity in 3D-IMC makes weight mapping a non-trivial problem, as weights must be placed such that those required by concurrent compute operations can be activated simultaneously, thereby maximizing resource utilization.

\section{Motivational study: Impact of weight mapping in 3D-IMC on runtime}

This work is motivated by the following insights into how the process of mapping an individual \ac{mvm} into the \ac{3dimc} macro impacts global throughput, and how the \ac{llm} inference context directly impacts the optimal choice of mapping.

% The mapping of the weights should be organized to maximize the utilization of the limited compute resources imposed by the \ac{ott} constraint and interconnect topology.
% The \ac{ott} constraint necessitates thoughtful mapping of weight matrices into the memory stack to make full use of the limited planar compute space.
% Figure~\ref{fig:active_weights} illustrates how a naive mapping results in underutilization of the compute space, as \acp{mvm} mapped to the same planar region on different tiers cannot operate simultaneously.
% This work proposes context-aware weight mapping to the 3D memory stack to improve the compute utilization of \ac{3dimc}, thereby improving \ac{mvm} and end-to-end latency and throughput.

\subsection{Mapping weights to 3D-IMC tiles and tiers}
\label{subsec:tiling}
The weights of \acp{mvm} are tiled before being mapped to \ac{3dimc} tiles.
The tiling and activation-flow strategy is similar to that of~\cite{jang2025}.
% Because the accumulators sit deep in the H-tree, each input vector must be broadcast to the LGs to which it is routed and then scattered internally.
As Figure~\ref{fig:architecture_dataflow}(f-g) illustrates, the weight matrix is first split column-wise along the output dimension and then row-wise along the input dimension.
The row split is chosen so that each block fills one leaf (four tiles).

Once an \ac{mvm}'s weights are tiled column- and row-wise, the tiled blocks in each column are placed on \acp{lg} as shown in Figure~\ref{fig:architecture_dataflow}(h).
All blocks of a column (shown with the same color in the figure) must reside on the same \ac{lg}, since accumulation can only occur within an \ac{lg}.
Crucially, within an \ac{lg}, the blocks of a column (different shades of the same color) can be mapped onto one or more leaves, e.g. 1, 2, or 4 leaves (or 4, 8, or 16 tiles) in macro configurations with 4 leaves per \ac{lg}.
We refer to the $n$-leaf option as $n$-way \ac{tp}, because the number of mapped leaves determines the degree of \ac{tp} for the corresponding \ac{mvm}.
Under the 1-way \ac{tp} option, all blocks are vertically stacked on the same four tiles in a leaf.
The 2- and 4-way \ac{tp} options spread each block across more tiles, reducing vertical stacking along a column.
This flexibility enables trade-offs between vertical and planar mappings, as discussed in Section~\ref{sec:t_p_tradeoff}.

At runtime, the input activation is broadcast to the \acp{lg} holding the weights of the corresponding \ac{mvm}.
Within each \ac{lg}, the input vector is scattered into chunks and routed to the relevant leaves depending on the chosen weight placement option.
The resulting output vectors are accumulated within a leaf and across leaves for 2- and 4-leaf \ac{tp}.
Finally, the per-\ac{lg} output vectors are transferred to the global IO buffer and concatenated.
Figure~\ref{fig:architecture_dataflow}(h-i) visualizes the process.

% \begin{figure}[htbp]
%     \centering
%     \includegraphics[width = \columnwidth]{figures/mapping_comparison.pdf}
%     \caption{
%         Visualizations of 2-/4-way \ac{tp} weight-mapping layouts of a Llama 70B transformer block on the \ac{imc} macro defined in Figure~\ref{fig:architecture_dataflow}.
%         Voxels correspond to an \acp{lg} on a tier.
%         2-way TP enables greater \ac{bp}, whereas 4-way \ac{tp} reduces single-\ac{mvm} latency.}
%     \label{fig:mapping_layouts}
% \end{figure}

\subsection{Single-MVM latency/batch-parallelism trade-off}
\label{sec:t_p_tradeoff}

% Figure~\ref{fig:architecture_dataflow}(h) shows three placement options for the row-wise-split blocks of the weight matrix.
% The 1-leaf option stacks all blocks onto the vertical tiers of four tiles in a single leaf.
% The 2-leaf option instead occupies eight tiles, spreading the blocks across two leaves.
% The 4-leaf option is the most aggressive placement, occupying all 16 tiles in an \ac{lg} with minimal vertical stacking.

% Under the \ac{ott} constraint, the 1-way \ac{tp} option fully serializes the \ac{mvm} across weight blocks due to the vertical stacking. 
% In contrast, choosing the 4-way \ac{tp} option for an \ac{mvm} parallelizes the \ac{mvm} across the tiled weight blocks, but at the cost of occupying more tiles and leaving less planar tile space for other \acp{mvm} on the \ac{imc} macro. 
A high degree of \ac{tp} parallelizes the \ac{mvm} across the tiled weight blocks by spreading the computation over planar resources, whereas lower \ac{tp} options make more use of the vertical space. This choice has direct performance implications. As an example, Figure~\ref{fig:intro}a visualizes the difference between 2-way and 4-way \ac{tp} layouts. The vertical z-axis is the tier axis, while the $8\times8$ horizontal plane corresponds to the \acp{lg}. The 2-way \ac{tp} option is marked as BP-friendly because it gives up a degree of \ac{tp} to enable more \ac{bp}: the \ac{ffn} \ac{g+u} \ac{mvm} (yellow voxels) can overlap with the output-projection and \ac{ffn} down-projection \acp{mvm} (green and purple voxels) across tokens, since they share no \acp{lg}. The advantage of 4-way \ac{tp} is that it uses fewer vertical tiers per \ac{mvm}, so less of the computation is serialized, giving shorter per-\ac{mvm} latencies. 
% The difference between 2-way and 4-way \ac{tp} layouts is visualized in Figure~\ref{fig:intro}a.
% The vertical z-axis is the tier axis, while the $8\times8$ horizontal plane corresponds to the \acp{lg}.
% Individual \ac{mvm} latency is lower with 4-way \ac{tp} because the computation of the \ac{mvm} spreads across more \acp{lg}; under 2-way \ac{tp}, however, the \ac{ffn} \ac{g+u} \ac{mvm} (yellow voxels) can overlap with the output-projection and \ac{ffn} down-projection \acp{mvm} (green and purple voxels) for other tokens because they share no \acp{lg}.
% This creates a trade-off between \ac{tp} and \ac{bp} at batch sizes greater than one.
The most suitable option should be chosen based on the targeted workload and expected runtime context.

\subsection{Impact of inference context on TP degree}

We find that the optimal selection of \ac{tp} degree for individual \acp{mvm} is heavily influenced by the \textit{inference context}, which can be broken down into three categories:

\textbf{Architectural characteristics}
such as the tier count and planar space define available height and width for \ac{mvm} shapes, while \ac{lg} size defines valid \ac{tp} degrees.
Similarly, a low-bandwidth or coarse-grained off-chip interconnect between the \ac{3dimc} accelerator and the attention unit may induce synchronization between tokens in a batch before communication, limiting \ac{bp} across different transformer layers.
Consequently, favoring \ac{bp} for weight mapping across layers provides little benefit, as the resulting computational parallelism cannot be effectively exploited.

% Similarly, a coarse-grained off-chip interconnect between the \ac{imc} chip and attention device may prevent \acp{mvm} from different layers from participating in \ac{bp}.
% These characteristics affect mapping in different ways.
% For example, memory organization determines weight-mapping granularity and therefore the extent to which \acp{mvm} can occupy disjoint resources.
% Similarly, a coarse-grained off-chip interconnect between the \ac{imc} chip and attention device may prevent \acp{mvm} from different layers from participating in \ac{bp}.

\textbf{Network characteristics} such as network type and size also affect the trade-off between \ac{tp} and \ac{bp}, as larger networks place greater pressure on limited compute resources.

Deserving special mention here are \Ac{moe} models, efficient alternatives to dense \acp{llm} as input-specific expert routing enables sparse \ac{ffn} computation~\cite{shazeer2017}.
\ac{3dimc} devices, with their skewed storage-to-compute capacity, naturally exploit \ac{moe} sparsity~\cite{Buchel2025}.

Mapping \ac{moe} \acp{llm} onto \ac{3dimc} accelerators raises two challenges.
First, \acp{moe} introduce the concept of \acf{ep}, where sibling experts (expert alternatives of the same \ac{ffn}) can be mapped such that they can run in mutually exclusive sets of IMC tiles in parallel. 
However, enabling \ac{ep} comes at the expense of \ac{tp} and \ac{bp}, further complicating the optimization space.
Second, the aggregate expert weights are larger than their dense counterparts even though only the top-$k$ experts are activated per layer.
The compute constraints of \ac{3dimc} make it infeasible to place all sibling experts such that they execute concurrently.
Mapping must therefore decide which experts to parallelize and which to serialize — a decision that can exploit domain-specific expert co-activation imbalances.
The extent of these imbalances is discussed in Appendix~\ref{sec:expert_imbalance}.

% First, the aggregate expert weights are larger than their dense counterparts even though only the top-$k$ experts are activated per layer.
% The compute constraints of \ac{3dimc} make it infeasible to place all experts of a single \ac{mvm} such that they execute concurrently.
% Mapping must therefore decide which experts to parallelize and which to serialize — a decision that can exploit domain-specific expert co-activation imbalances.
% The extent of these imbalances are discussed in Appendix~\ref{sec:expert_imbalance}.
% Second, \acp{moe} introduce the concept of \acf{ep}, where sibling experts can be mapped parallel to improve latency, even in single-batch scenarios, at the expense of \ac{tp} and \ac{bp}, further complicating the optimization space.

\textbf{Runtime characteristics} such as expected batch size and sequence length, further impact mapping.
Single-batch, short-context inference is typical at the edge~\cite{chen2026}, where minimizing latency through high \ac{tp} is preferred.
In contrast, data centers typically serve large batches over long sequences, where maximizing batch-level overlap to achieve high throughput is more desirable.

% When the expected inference context is known, the deployment and inference stacks can be tuned accordingly, subject to power-budget and \ac{slo} trade-offs~\cite{lin2026}.
% Single-batch, short-context inference is typical at the edge~\cite{chen2026}, where minimizing latency through high \ac{tp} is preferred.
% In contrast, data centers typically serve large batches over long sequences, where maximizing batch-level overlap to achieve high throughput is more desirable.
% On \ac{3dimc} devices, weight mapping is among the most consequential deployment decisions~\cite{houshmand2024} for tuning performance to the expected inference regime.

\subsection{Underutilization of the 3D-IMC accelerator}

\begin{figure}[t]
    \centering
    \includegraphics[width = 0.9\columnwidth]{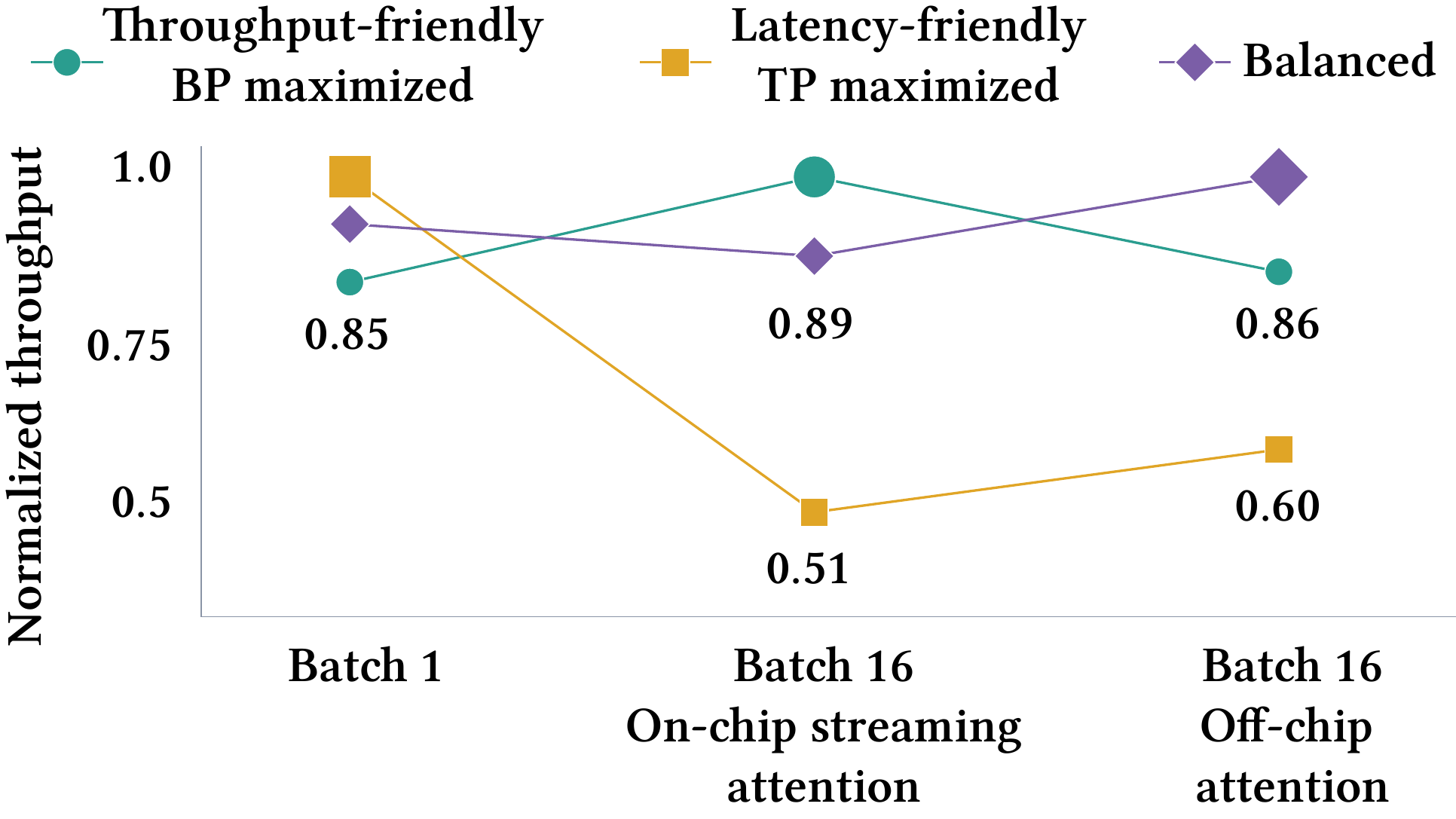}
    \Description{Line chart of normalized throughput for throughput-friendly, latency-friendly, and balanced mappings under three serving contexts.
    At batch size 1, the latency-friendly mapping is best at 1.0 while the other mappings are near 0.85--0.9.
    At batch size 16 with on-chip streaming attention, the throughput-friendly mapping is best at about 1.0 and the latency-friendly mapping falls to 0.51.
    With off-chip attention at batch size 16, the balanced mapping is best at about 1.0, followed by the throughput-friendly mapping at 0.86 and the latency-friendly mapping at 0.60.}
    \caption{Impact of \acs*{3dimc} mapping strategies on decode-step throughput for Llama 3.1-70B across representative runtime contexts.
    Results shown for a sequence length of 128.}
    \label{fig:mapping_motive}
\end{figure}

Existing \ac{3dimc} mapping algorithms~\cite{Buchel2025, bavandpour2019} are context-agnostic and maximize \ac{tp}, underutilizing compute resources even at saturating batch sizes.
Figure~\ref{fig:intro}e shows that the context-agnostic method presented in~\cite{Buchel2025} utilizes only $42\%$ of the \ac{3dimc} compute capacity on average.
Its objective minimizes the latency of each \ac{mvm} in isolation and therefore provides no incentive to enable concurrent \ac{mvm} execution across tokens in a batch.

Figure~\ref{fig:mapping_motive} illustrates the impact of weight mapping across inference contexts using three alternatives: a throughput-friendly mapping that promotes overlapped execution across tokens in a batch, a latency-friendly mapping that maximizes \ac{tp} to minimize latency, and a balanced mapping.
The throughput-friendly mapping improves throughput by up to $1.96\times$ at a saturating batch size of 16, whereas the latency-friendly mapping is preferable in lower-batch contexts.

Orthogonally, we further vary the inference context by changing the attention granularity.
If attention is conducted per token on-chip in a \textit{streamed} fashion, it does not prevent tokens from executing in parallel across different transformer layers.
Then throughput-friendly mapping is preferable with potentially high~\ac{bp}.
% When attention is incorporated as an additional overlappable stage during inference in a \textit{streamed attention} fashion, the throughput-friendly mapping is preferable.
In contrast, when attention is disaggregated off-chip and operates in a batched fashion such that static \acp{mvm} and attention cannot overlap across tokens, the more balanced mapping outperforms the throughput-friendly mapping.
Thus, because the optimal layout depends on the inference context, a context-agnostic mapping strategy~\cite{Buchel2025,huang2025,bavandpour2019} may not fully realize the computational potential of the system.
Our work targets precisely this gap.

\section{CAMP Mapping Algorithm}
\label{sec:mapping_algorithm}

\begin{figure}[tbp]
    \centering
    \includegraphics[width=0.90\columnwidth]{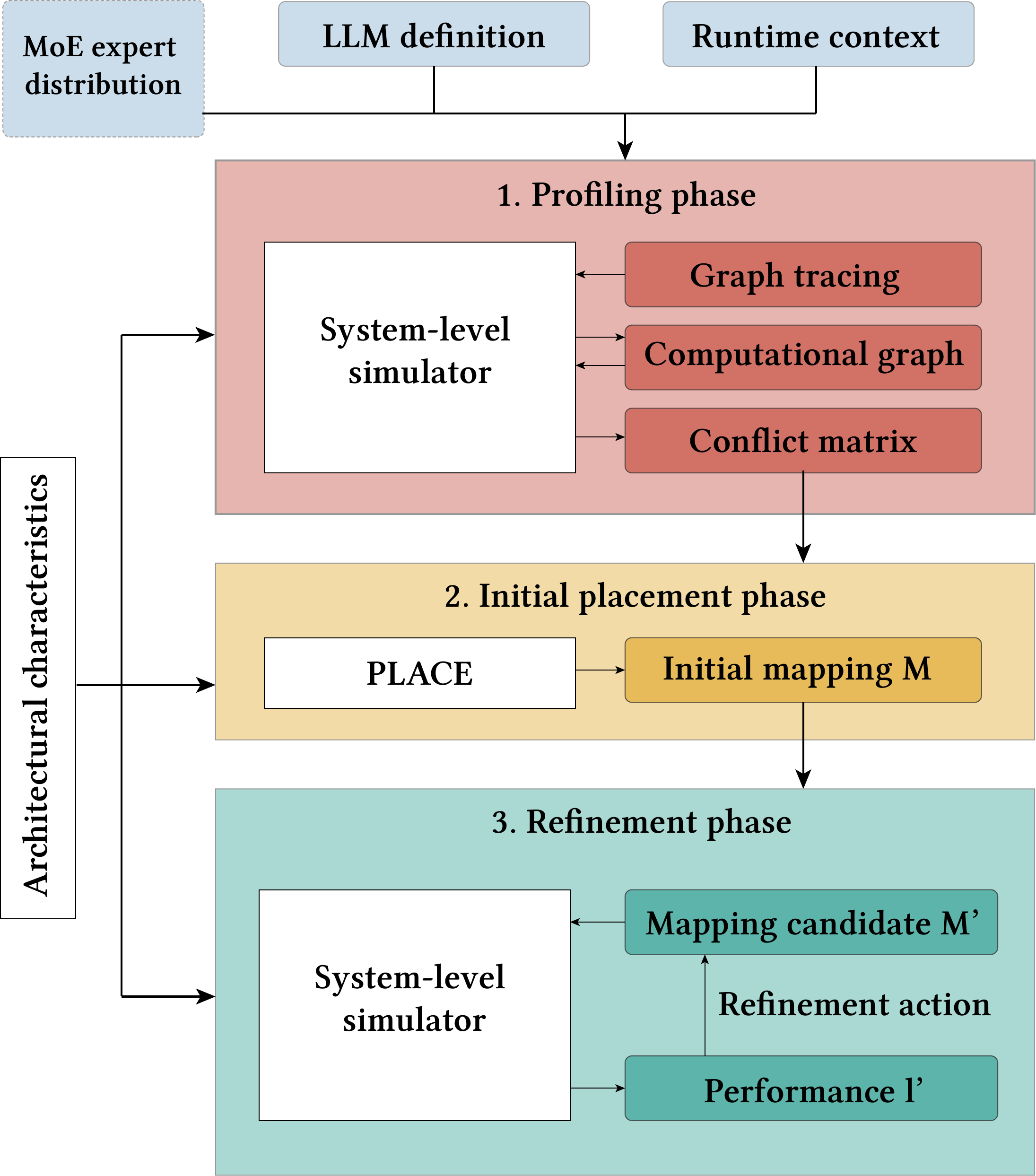}
    \Description{Flowchart of \acs*{camp}'s three phases.
    The \acs*{llm} definition, runtime context, optional mixture-of-experts distribution, and system configuration feed a profiling phase in which graph tracing and a computational graph are evaluated by a system-level simulator to produce a conflict matrix.
    The initial-placement phase applies Place to produce mapping M.
    The refinement phase simulates candidate mapping M prime, measures performance l prime, applies another refinement action, and feeds the candidate back until the best mapping is selected.}
    \caption{Overview of the \acs*{camp} mapping algorithm.}
    \label{fig:alg_overview}
\end{figure}

To exploit this mapping flexibility and maximize \ac{3dimc} throughput, \ac{camp} uses the three-stage flow illustrated in Figure~\ref{fig:alg_overview}.
\ac{camp} comprises \textit{profiling}, \textit{initial-placement}, and \textit{refinement} phases, described in the following subsections.
% In the \textit{profiling phase}, we capture the input workload characteristics and encode them into a conflict matrix.
% In the \textit{initial placement phase}, we establish an initial weight mapping.
% Finally, in the \textit{refinement phase}, we iteratively optimize the mapping through successive refinements.
% The following subsections describe each of these three phases in detail.

\begin{figure*}[htbp]
    \centering
    \includegraphics[width=\textwidth,trim={0.0cm 1cm 0cm 0cm},clip]{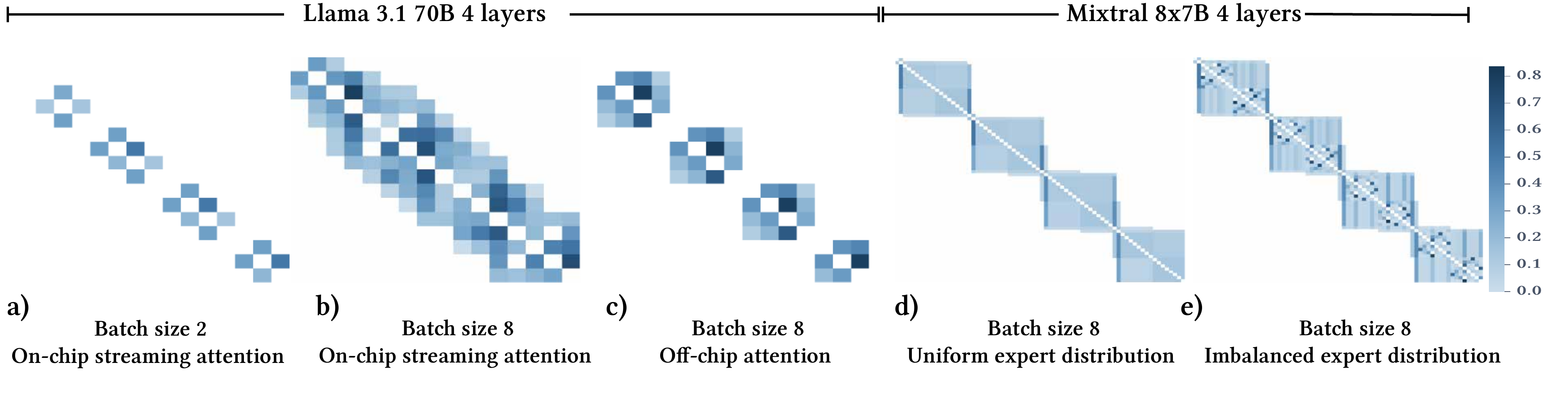}
    \Description{Five blue heatmaps whose intensity encodes pairwise \acs*{mvm} temporal proximity from zero to above 0.8.
    For four Llama 3.1 70B layers, panel (a), batch size 2 with on-chip streaming attention, contains small sparse diagonal clusters; panel (b), batch size 8 with streaming attention, becomes a broad connected diagonal band; and panel (c), batch size 8 with off-chip attention, separates into four isolated clusters.
    For four Mixtral 8-by-7B layers at batch size 8, panel (d) shows broad block-diagonal conflicts under a uniform expert distribution, whereas panel (e) shows nonuniform stripes and spots within the blocks under an imbalanced distribution.}
    \caption{Conflict matrices shown as heatmaps across workload configurations and inference contexts.
             An entry in row $i$ and column $j$ corresponds to the temporal proximity of \acsp*{mvm} $i$ and $j$ among the set of \acsp*{mvm} to be mapped.}
    \label{fig:conflict_matrices}
\end{figure*}

\subsection{Profiling phase}

% \subsubsection{The conflict matrix $\mathbf{C}$ for conventional LLMs}
% \label{sec:conflict_matrix}

The profiling phase quantifies the \textit{temporal proximity} between \acp{mvm} while accounting for the target architecture, model, and runtime context.
This information is subsequently used to identify \acp{mvm} that are likely to execute concurrently and thus should occupy disjoint sets of \acp{lg}, enabling \ac{bp}.

% To obtain this information, we first extract the set of \acp{mvm} and their tensor dimensions from the target LLM under the serving regime of interest.
First, we trace model execution under the expected runtime context and lower the resulting computation to a graph-based \ac{ir}.
The \ac{ir} is then parsed to identify the dimensions of each \ac{mvm} mapped to the \ac{3dimc} macro and the estimated latencies of other operations, including layer normalization, activation functions, and attention.
A system-level simulator then schedules the computational graph to extract network runtime metrics.
During this profiling phase, we relax the physical resource constraints by providing enough tiles to map every \ac{mvm} with 4-way \ac{tp} onto mutually exclusive \acp{lg}.
% During this profiling phase, we relax the physical system constraints by assuming an unlimited number of leaf groups, i.e., $N_{LG} \rightarrow \infty$, rather than the number of leaf groups in the target configuration ($N_{LG}=64$).
This configuration allows all independent \acp{mvm} to execute concurrently, exposing their maximum potential temporal overlap, without contention for compute resources.
% Although larger than the target macro, it isolates the \acp{mvm}' intrinsic temporal proximity from contention caused by limited \acp{lg}, leaves, and tiles.

The simulation provides a schedule of operations consisting of their start and end times.
Let $\mathcal{J}$ denote the set of \acp{mvm} in the workload.
For each \ac{mvm} $j\in\mathcal{J}$ and each token $b \in \{1,\ldots,B\}$, where $B$ is the batch size, the schedule defines start and end times $s_{j,b}$ and $e_{j,b}$.

The schedule satisfies two constraints:
(i) two tokens cannot execute the same \ac{mvm} concurrently, since they contend for the same physical array;
(ii) data dependencies prevent the \acp{mvm} of a single token from executing concurrently, except for the expert layers of \ac{moe}-based networks.

\begin{equation}
\begin{aligned}
s_{j,b+1} \geq e_{j,b},
\quad
\forall j\in\mathcal{J},\; 1 \leq b < B \\
s_{j+1,b} \geq e_{j,b},
\quad
\forall b \leq B
\end{aligned}
\label{eq:schedule_constraints}
\end{equation}

% \begin{equation}
% s_{j,b} \geq e_{i,b},
% \qquad
% \forall b \leq B.
% \label{eq:disjoint_kernels}
% \end{equation}

% Data dependencies also prevent \acp{mvm} in the same batch from executing concurrently, except within expert layers in MoE-based networks. %when they are ordered by the dependencies of the computation graph.
% More generally, for \acp{mvm} $i$ and $j$ where $j$ depends the output of $i$:

Given this schedule, we quantify the temporal overlap $O_{i,j}$ between two \acp{mvm} $i$ and $j$ by summing the overlap of all pairs of their token instances:

\begin{equation}
O_{i,j} = \sum_{b=1}^{B} \sum_{b'=1}^{B}
\max\!\left(0,\, \min(e_{i,b}, e_{j,b'}) - \max(s_{i,b}, s_{j,b'}) \right)
\label{eq:overlap}
\end{equation}

The \textit{temporal proximity} of $i$ to $j$, $C_{i,j}$, is defined as:
%the fraction of $i$'s total execution time that overlaps with instances of $j$:

\begin{equation}
C_{i,j} =
\dfrac{O_{i,j}}{\sum_{b=1}^{B} \left( e_{i,b} - s_{i,b} \right)}
\label{eq:conflict}
\end{equation}

Thus, $C_{i,j}\in[0,1]$ represents the portion of \ac{mvm} $i$'s execution that potentially overlaps with \ac{mvm} $j$.
% This definition is directional: $C_{i,j}$ need not equal $C_{j,i}$ because the entries are normalized by the total execution times of different \acp{mvm}.
Scheduling constraints (Equation~\ref{eq:schedule_constraints}) ensure that no overlap is double-counted.
% Within an \ac{mvm}, batch-instance intervals are disjoint by Equation~\ref{eq:sc}, while within a batch, graph dependencies prevent dependent \acp{mvm} from overlapping by Equation~\ref{eq:disjoint_kernels}.
The resulting \textit{conflict matrix} is:

\begin{equation}
    \mathbf{C} \in [0,1]^{\mathcal{J} \times \mathcal{J}},
\quad\mathbf{C}_{i,j} =
\begin{cases}
0 & i = j \wedge \text{LLM is dense} \\
C_{i,j} & \text{otherwise}
\end{cases}
\end{equation}

For dense \acp{llm}, diagonal entries are zero because an \ac{mvm} cannot conflict with itself during placement.
For \ac{moe}-based \acp{llm}, however, multiple expert instances of the corresponding dense \ac{mvm} can execute concurrently, requiring the different treatment described below.

\subsubsection{Expanding the conflict matrix for MoE LLMs}

For dense \acp{llm}, the set of \acp{mvm} $\mathcal{J}$ contains the \ac{qkv}, output, \ac{g+u}, and down projections of each transformer layer.
In an \ac{moe}-based model, each \ac{g+u} and down projection is instead replicated across $E$ expert \acp{mvm}, of which $K$ are activated per token.

Representing expert-level conflicts introduces two considerations.
% The conflict matrix obtained using the procedure in Section~\ref{sec:conflict_matrix} does not explicitly represent these expert-level conflicts.
%The expert decomposition introduces two additional considerations.
First, multiple experts in the same \ac{ffn} block are activated concurrently.
Thus, the self-conflicts (diagonal entries in the conflict matrix) corresponding to \ac{ffn} blocks are non-zero, unlike those of dense \ac{ffn} blocks.
%An expert \ac{ffn} block can therefore have a nonzero self-conflict when its constituent experts occupy different \acp{lg}, unlike a conventional \ac{ffn} block.
Second, expert activation is not perfectly uniform: some expert subsets are activated more frequently than others.
% Representing an entire expert \ac{ffn} block as one \ac{mvm} obscures these differences and prevents \ac{camp} from exploiting them.
To model expert-level conflicts, we decompose the \ac{ffn} \acp{mvm} into individual expert \acp{mvm}.
% We address this limitation by relating the expert network to its dense counterpart and expanding the dense set of \acp{mvm} to represent each expert \ac{mvm} individually.
Let $\mathcal{J}'$ denote the expanded set of \acp{mvm}.
For an expert \ac{mvm} $i$, $\mathcal{J}'$ contains one entry $(i,e)$ for each expert $e\in\mathcal{E}$, whereas a non-expert \ac{mvm} $i$ remains represented by a single entry $i$.
We define a mapping from $\mathcal{J}'$ to $\mathcal{J}$:
\begin{equation}
\pi(i,e)=i, \quad \pi:\mathcal{J}'\rightarrow\mathcal{J}
\end{equation}

This mapping associates each expanded \ac{mvm} with its corresponding \ac{mvm} in the dense network. $\pi(i)=i$ for a non-expert \ac{mvm}.
We expand the dense conflict matrix $\mathbf{C}$ into the more expressive $\widetilde{\mathbf{C}} \in [0,1]^{|\mathcal{J'}| \times |\mathcal{J'}|}$:

\begin{equation}
    \widetilde{\mathbf{C}}_{k,l} = \mathbf{C}_{\pi(k),\pi(l)} \cdot \frac{\omega(k,l)}{\sum_{l' \in \text{siblings}(l)} \omega(k,l')} \quad \forall (k,l)\in |\mathcal{J'}| \times |\mathcal{J'}|
    \label{eq:moe_conflict_expansion}
\end{equation}

The weighting function $\omega$ distributes each dense-\ac{mvm} conflict among the corresponding expert pairs in proportion to their use.
Equation~\ref{eq:moe_conflict_expansion} conserves each row: for any $k$ and original \ac{mvm} $j$, the expanded entries sum to the original conflict, $\sum_{l:\pi(l)=j}\widetilde{\mathbf{C}}_{k,l}=\mathbf{C}_{\pi(k),j}$.
For a perfectly balanced \ac{moe} network, $\omega$ is constant and distributes conflicts uniformly across experts.
Appendix~\ref{sec:expert_imbalance} defines $\omega$ for non-uniform expert-activation distributions.
For simplicity, we write $\mathbf{C}$ and $\mathcal{J}$ throughout the remainder
of the paper; for \ac{moe} models these denote the expanded $\widetilde{\mathbf{C}}$
and $\mathcal{J}'$.

\begin{algorithm}[t]
\caption{Initial placement phase and placement algorithm}
\label{alg:initial-placement}
{\small
\begin{algorithmic}[1]
\Require MVMs $\mathcal{J}$ in computation order; conflict matrix $\mathbf{C}$;
         leaf groups $\mathcal{H}$; tier budget $T$
\Ensure  mapping $\mathbf{M}$: $\mathbf{M}_{j,h,t} \in \{0,1\}$ whether MVM $j$ occupies leaf group $h$ in tier $t$
\State $\text{TP}_j \gets 1 \quad \forall j \in \mathcal{J} $ ;
\State \Return $\textsc{Place}(\mathcal{J})$
%\Statex
\Function{Place}{$\mathcal{J}$}
\State $\mathbf{M}_{j,h,t} \gets 0 \quad \forall j\in\mathcal{J}, h \in \mathcal{H}, t < T$
\ForAll{$j \in \mathcal{J}$}
  \State $(h_j, \tau_j) \gets \textsc{Footprint}(j, \text{TP}_j)$
  \State $\mathcal{P} \gets \textsc{Candidates}(\mathbf{M}, h_j, \tau_j)$
  \ForAll{$\textbf{M'} \in \mathcal{P}$}
    \State $u_{j,h} \gets \underset{t<T}{\max} \hspace{5pt} \mathbf{M'}_{j,h,t} \quad \forall j\in\mathcal{J}, h \in \mathcal{H}$
    \State $x_{i,j} \gets \mathbf{1}\big[\big(\sum_{h \in \mathcal{H}} u_{i,h} \cdot u_{j,h}\big) \geq 1 \big]$
    \State $\gamma(\textbf{M'}) \gets \sum_{i<j} \mathbf{C}_{i,j} \cdot x_{i,j} \quad \forall i,j\in\mathcal{J}$
  \EndFor
  \State $\mathbf{M} \gets \arg\min_{\mathbf{M'} \in \mathcal{P}}
         \big(\, \gamma(\mathbf{M'})\big)$
\EndFor
\State \Return $\mathbf{M} $
\EndFunction

\end{algorithmic}
}
\end{algorithm}

\subsubsection{Inference context encoded by the conflict matrix}

The conflict matrix captures all three categories of inference context, as illustrated by the heatmaps in Figure~\ref{fig:conflict_matrices}.

\noindent\textbf{Architectural characteristics:} Figure~\ref{fig:conflict_matrices}(b,c) compares conflict matrices for the settings with on-chip and off-chip attention units, respectively.
When the attention unit is integrated on-chip, as in~\cite{Buchel2025,yu2024}, individual batch tokens can be streamed into the attention unit at fine granularity, enabling continuous inter-layer overlap across tokens in the batch.
This fine-grained overlap produces the connected conflict matrix in Figure~\ref{fig:conflict_matrices}(b).
When the attention unit is disaggregated off-chip~\cite{xu2026}, gathering the accelerator outputs for off-chip communication introduces a barrier between layers, preventing overlap between the preceding \ac{qkv} projection layer and the subsequent output projection layer.
This barrier produces the disconnected conflict matrix in Figure~\ref{fig:conflict_matrices}(c).

\noindent\textbf{Network characteristics:} Figure~\ref{fig:conflict_matrices}(a-c) represents Llama 70B, whereas Figure~\ref{fig:conflict_matrices}(d,e) represents Mixtral 8$\times$7B with expert layers decomposed into individual experts.
Incorporating the expected expert-activation distribution allows \ac{camp} to account for co-activation likelihood.

\noindent\textbf{Runtime characteristics:} Figure~\ref{fig:conflict_matrices}(a,b) illustrates how batch size affects the conflict matrix.
Smaller batch sizes produce fewer potential conflicts between \ac{mvm} pairs and therefore sparser conflict matrices. 

Note that the \textbf{conflict matrix is just an intermediate result}, guiding the rest of the algorithm when mapping the weights. The next two phases run in a workload-agnostic manner as all the workload-related characteristics are encoded into the conflict matrix.

\subsection{Initial placement phase and placement algorithm}

%While the profiling phase serves as preparation, 
The initial placement phase produces the first weight mapping from the profiling results.
Algorithm~\ref{alg:initial-placement} describes the placement process.
Because the \textsc{Place} function is greedy and order dependent, \acp{mvm} are first arranged in computation order, following the heuristic in~\cite{Buchel2025}.

Once ordered, \acp{mvm} are placed consecutively.
For each \ac{mvm} $j$ and its \ac{tp} option, $\text{TP}_j$, we first compute its storage footprint in terms of the number of \acp{lg} ($h_j$) and vertical tiers ($\tau_j$) required.
We then enumerate all combinations of at least $h_j$ \acp{lg} that can be targeted by a single H-tree broadcast step, together with their first $\tau_j$ consecutive empty tiers, to form the set of candidate mappings.
Each candidate is scored by a scoring function $\gamma$, using the conflict matrix $\mathbf{C}$ to quantify the weighted conflicts between the \acp{mvm} already placed and the candidate mapping.
The candidate with the minimum total weighted conflict is picked.
%as the placement for the current \ac{mvm}.
This process is repeated until all \acp{mvm} are mapped.

During initial placement, we assign a 1-way \ac{tp} option to every \ac{mvm}.
This minimizes the number of tiles, leaves, and \acp{lg} required by each \ac{mvm}, at the cost of occupying more vertical tiers.
This configuration serves as a starting point for the refinement phase.

\begin{algorithm}[t]
\caption{Refinement phase}
\label{alg:refinement}
{\small
\begin{algorithmic}[1]
\Require initial mapping $\mathbf{M}_{\mathrm{init}}$;
         beam width $W$; action width $k$
\Ensure refined mapping $\mathbf{M}$

\State $l_{\mathrm{init}} \gets \textsc{Schedule}(\mathbf{M}_{\mathrm{init}})$
\State $S_{\mathrm{init}} \gets (\mathbf{M}_{\mathrm{init}}, l_{\mathrm{init}})$; $B \gets \{S_{\mathrm{init}}\}$; $S^* \gets S_{\mathrm{init}}$;

\While{$B \neq \emptyset$}
    \State $B_{\mathrm{cand}} \gets \emptyset$

    \ForAll{$S=(\mathbf{M},l) \in B$}
        \State $A(S) \gets
        \{a_{j,\text{TP}_{j,S}\rightarrow \text{TP}_{\mathrm{new}}}
        \mid j\in\mathbf{\mathcal{J}},
        \text{TP}_{\mathrm{new}}>\text{TP}_{j,S}\}$

        \State $A_k(S) \gets$ top $k$ actions in $A(S)$

        \ForAll{$a \in A_k(S)$}
            \State $\mathbf{M}' \gets \textsc{Place}(a(\mathbf{M}))$
            \State $l' \gets \textsc{Schedule}(\mathbf{M}')$
            \State $B_{\mathrm{cand}} \gets
            B_{\mathrm{cand}} \cup \{(\mathbf{M}',l')\}$
        \EndFor
    \EndFor

    \State $S^* \gets \arg\min_{(\mathbf{M}',l')\in
B_{\mathrm{cand}}\cup\{S^*\}} l'$
    \State $B \gets$ top $W$ states in $B_{\mathrm{cand}}$
\EndWhile

\State \Return $\mathbf{M}^*$ from $S^*$
\end{algorithmic}
}
\end{algorithm}

\subsection{Refinement phase}
\label{sec:refinement}

The refinement phase iteratively improves mapping by modifying the \ac{tp} options of individual \acp{mvm}.
\ac{camp} explores the search space of \ac{tp} degree combinations using a \textbf{beam search} with the system-level simulator in the feedback loop.
Algorithm~\ref{alg:refinement} presents the refinement pseudocode.

Initially, all \acp{mvm} are mapped using 1-way \ac{tp}, maximizing \ac{bp} at the cost of \ac{tp}.
The refinement phase determines whether increasing the \ac{tp} degree of selected \acp{mvm} provides a favorable trade-off between lower individual \ac{mvm} latency and compromised \ac{bp}.
% When the trade-off is favorable, \ac{camp} retains the higher \ac{tp} option and remaps the \acp{mvm}.
We call such a revision an \textit{action}: $a_{j,m \rightarrow n}$ changes \ac{mvm} $j$ from $m$-way to $n$-way \ac{tp}.
Applying an action yields a new mapping, and the pair of a mapping $\mathbf{M}$ and its latency $l$ forms a \textit{state} $S = (\mathbf{M}, l)$.
Since evaluating every assignment is intractable, \ac{camp} keeps only the $W$ most promising states after each iteration, a \textit{beam} of width $W$.
The search begins from a single state, the initial placement, with candidate actions:
\[
A = \{a_{j,1\rightarrow2} \mid j \in \mathcal{J}\}.
\]

Each iteration proceeds as follows.
For every state in the beam, \ac{camp} orders the candidate actions by the compute latency of their corresponding \acp{mvm} under that state's \ac{tp} degrees, prioritizing higher-latency \acp{mvm},
and expands the top $k$.
Expanding an action updates the \ac{tp} option of its \ac{mvm} and remaps the weights of \textit{all} \acp{mvm} using $\textsc{Place}$ from Algorithm~\ref{alg:initial-placement}.
The system-level simulator then evaluates
the resulting mapping $\mathbf{M}'$ and yields its latency $l'$, producing a new state $S' = (\mathbf{M}', l')$.
Of the $W\times k$ states generated this way, the $W$ with the lowest latency form the next beam.
When no candidate actions remain, \ac{camp} returns the mapping $\mathbf{M}^*$ of the lowest-latency state explored.

% We call the pair of a mapping $\textbf{M}$ and the resulting latency $l$ a \textit{state}, denoted by $S$.
% Each iteration of the beam search stores a \textit{beam} of states.
% There are $W$ states in a beam and $W$ is called the beam width.
% We then refer to a change in a \ac{mvm}'s \ac{tp} option as an \textit{action}.
% Specifically, action $a_{j,m\rightarrow n}$ denotes changing \ac{mvm} $j$ from $m$-way \ac{tp} to $n$-way \ac{tp}.
% At the first iteration, the set of candidate actions is:
% \[
% A = \{a_{j,1\rightarrow2} \mid \forall j \in \mathcal{J}\}.
% \]

% Actions are ordered by the compute latency of their corresponding \acp{mvm} under the current \ac{tp} options, prioritizing higher-latency \acp{mvm}.
% \ac{camp} explores the top $k$ actions.
% For each action, it updates the corresponding \ac{mvm}'s \ac{tp} option and remaps the weights of \textit{all} \acp{mvm} using $\textsc{Place}$ from Algorithm~\ref{alg:initial-placement}.
% % Unlike the initial placement phase, however, the \ac{tp} options of the \acp{mvm} vary during refinement.
% The system-level simulator evaluates the resulting mapping $\mathbf{M}'$ and yields latency $l'$, producing the refined state $S' = (\mathbf{M}', l')$.
% Each iteration retains the $W$ lowest-latency states in a beam.
% The next iteration repeats action selection, remapping, and simulation from these states.
% When no actions remain, \ac{camp} returns the mapping $\mathbf{M}^*$ from the lowest-latency explored state.
%Figure~\ref{fig:beam_search} illustrates this process.

As beam width $W$ and action width $k$ approach infinity, beam search approaches exhaustive exploration of the \ac{tp} configuration space.
In practice, their values balance search quality against simulation cost, thus depending on the number of \acp{mvm} and the cost of each simulation.
Appendix~\ref{sec:search_space_cardinality} analyzes the scale of the weight-mapping problem and shows that exhaustive exploration is infeasible at this scale, while Appendix~\ref{app:beam_search} provides ablations on $W$ and $k$.

%demonstrating that search quality saturates.

% It should be noted that exhaustive exploration is impractical because assigning each \ac{mvm} in $\mathcal{J}$ one of the 1-, 2-, or 4-way \ac{tp} options yields $3^{|\mathcal{J}|}$ possible configurations.
% For dense \acp{llm}, $|\mathcal{J}| = 4 \times$ the number of transformer layers.
% For \ac{moe}-based \acp{llm}, it grows to $(2 + 2E)$ times the number of layers, where $E$ is the number of experts.
% This complexity is further exacerbated by the exact placement options on \ac{lg} and tier combinations.
% Analysis on the search space cardinality for this problem can be found on Appendix~\ref{sec:search_space_cardinality}.
% Beam search provides a compromise between greedy and exhaustive search.

% \begin{figure}[htbp]
%     \centering
%     \includegraphics[width = 0.9\columnwidth]{figures/attention_ffn_v2.pdf}
%     \caption{Workload mapping with attention/\ac{mvm} disaggregation.}
%     \label{fig:attention_ffn}
% \end{figure}

\section{Experimental setup}
\label{sec:experimental_setup}

% We evaluate \ac{camp} across a range of architectures and networks to quantify its benefit over mapping algorithms in the literature.

\subsection{3D-IMC macro configuration}
%The \ac{imc} macro is comprised of tiles organized hierarchically into leaves and \acp{lg}.
We consider a 512 $\times$ 512, 320-tier \ac{aimc} tile~\cite{cho2025}, with 1,024 such tiles required to accommodate the parameter counts of SotA \acp{llm} such as Llama-3.1-70B and Mixtral-8$\times$7B.
The \ac{mvm} latency of each \ac{aimc} tile is 100\,ns, consistent with~\cite{Buchel2025}.
Leaves are interconnected by an H-tree architecture similar to the designs in~\cite{shim2021,jang2025}.
% Unlike those designs, this H-tree does not narrow at intersections because it must broadcast inputs to multiple \acp{lg} simultaneously.
Further, we parameterize the number of H-tree channels available for transmitting data.
Each channel has a width matched to the tile dimension, 512, and is deeply pipelined, allowing 64 8-bit inputs to be transmitted each cycle.
We use $n_{\text{LG}} = 64$ \acp{lg} with $n_L = 4$ leaves per group, and four H-tree channels.
Appendix~\ref{app:configsweep} presents the ablation supporting these choices and results for other macro configurations.

% Using the system-level designs presented in Refs.~\cite{shim2021,jang2025} as a baseline, the configuration details are adjusted to accommodate the target \acp{llm} and the hardware specifications of the adopted \ac{imc} tiles~\cite{cho2025,Buchel2025}.
% Specifically, we consider a 512 $\times$ 512, 320-tier \ac{imc} tile~\cite{cho2025}, with 1{,}024 such tiles required to accommodate the parameter counts of state-of-the-art \acp{llm} such as Llama-3.1-70B and Mixtral-8$\times$7B.
% The system is organized into an 8-level H-tree and arranged in a hierarchical topology of leaf groups and leaves.
% Unlike previous literature, the width of the H-tree is not reduced at intersections and is significantly more performant, as it is necessary to broadcast inputs to multiple leaf groups simultaneously.
% The width of the H-tree corridors is matched to the tile dimension, 512, and it is deeply pipelined, allowing 64 8-bit inputs to be transmitted each cycle.
% The number of leaf groups and leaves per leaf group is set to 64 and 4, respectively, based on an ablation study, the results of which are presented in Appendix~\ref{app:configsweep}.

\begin{figure*}[htbp]
    \centering
    \includegraphics[width=1.0\textwidth]{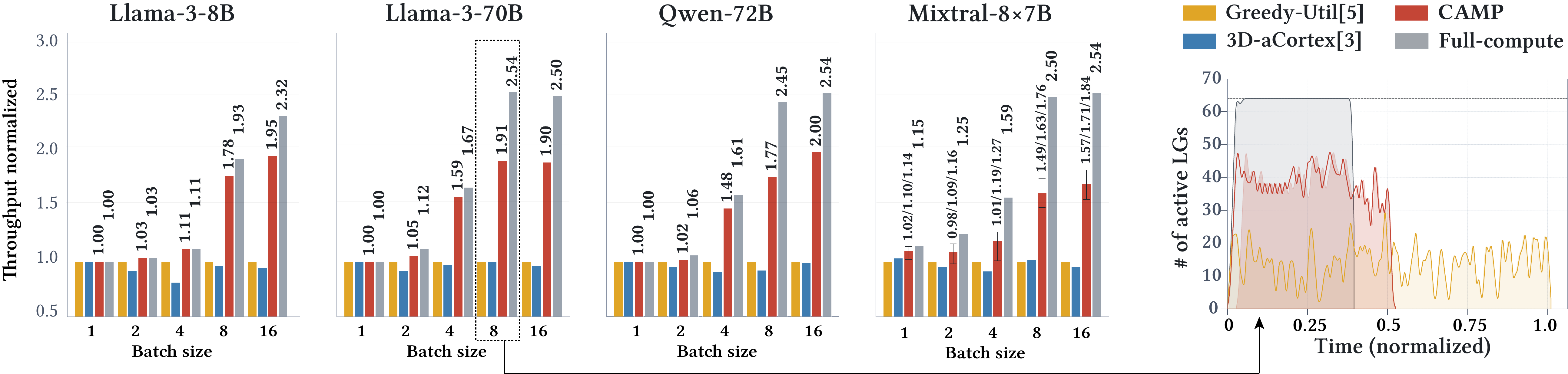}
    \Description{Four grouped-bar charts report throughput normalized to Greedy-Util for Llama 3 8B, Llama 3 70B, Qwen 72B, and Mixtral 8-by-7B at batch sizes 1, 2, 4, 8, and 16.
    \acs*{camp} matches the baselines near batch size 1, then rises with batch size to approximately 1.95, 1.90, 2.00, and 1.71 times throughput, respectively, while 3D-aCortex remains near or below one; the gray full-compute bounds are higher.
    A linked time trace for Llama 70B at batch size 8 shows \acs*{camp} sustaining roughly 35--48 active leaf groups for about half the normalized duration, whereas Greedy-Util usually activates fewer than 25 leaf groups and runs for the full duration.}
    \caption{\acs*{mvm} speedup across mapping algorithms, workloads, and batch sizes.
    An additional plot for ongoing compute utilization in terms of the number of active \acsp*{lg} is also appended.}
    \label{fig:imc_speedup}
\end{figure*}

\subsection{Attention/MVM disaggregation setup}
Our speedup evaluation targets the memory-bound decode phase.
Within decode, we apply attention/static \ac{mvm} disaggregation, assigning static \acp{mvm} to the \ac{3dimc} accelerator, while assigning attention to the attention unit(s) under varying configurations, such as 1 or 2 NVIDIA A100 \acp{gpu} (80GB), H100 \acp{gpu} (80GB), or B200 \acp{gpu} (180 GB).
%The workload mapping is visualized in Figure~\ref{fig:attention_ffn}.

\subsection{Target models and mapping configuration}
\label{sec:simulation}
We evaluate \ac{camp} on Llama-3.1-8B/70B~\cite{touvron2023}, Qwen-72B~\cite{bai2023}, and Mixtral-8$\times$7B~\cite{jiang2024}.
Results are obtained by combining \ac{gpu}-profiled attention benchmarks with a Torch-integrated \ac{3dimc} simulator.
The details on profiling \ac{gpu}-related metrics are provided in Appendix~\ref{sec:gpu_profiling}.

% Since transformer layers are repetitive within a network, mapping one layer's weights and replicating that mapping for the others is a feasible solution.
% However, replicating a single layer's mapping is not sufficient to enable inter-layer pipelining, nor to adapt to the varying expert distributions across \ac{moe} layers.
Ideally, mapping all layers at once with \ac{camp} would yield the most efficient solution.
The runtime of the proposed algorithm, however, scales quadratically with the number of \acp{mvm} being mapped.
A detailed complexity analysis is provided in Appendix~\ref{app:runtime_complexity}.
As a tradeoff between mapping runtime and inter-layer pipelining, we run \ac{camp} on eight layers for the dense networks and replicate the resulting mapping across the remaining layers.
For Mixtral, we run \ac{camp} on groups of four layers, solving for each group independently, as expert activation distributions differ across layers unless a uniform distribution is assumed.
We allocate the vertical tiers in proportion to the number of layers being mapped, so that the total number of occupied tiers remains feasible.
Mixtral results are averaged over $50$ trials to mitigate the randomness of expert activation, which corresponds to $1600 \times \text{batch size}$ expert-pair samples.

\subsection{Comparison baselines}
We compare \ac{camp} mapping results against two \ac{3dimc} weight-mapping algorithms from the literature:
\begin{enumerate}[leftmargin=*]
    \item \textbf{Greedy-Util}~\cite{Buchel2025}: The Greedy-Util benchmark greedily maps \acp{mvm} in execution order, maximizing \ac{tp} for all \acp{mvm}.
    The algorithm selects the set of accelerator tiles with the lowest tile utilization, computed by counting the number of non-empty tiers on each tile.
    For expert \acp{mvm}, this count is weighted by $1/E$.
    As a result, sibling expert \acp{mvm} are stacked onto the same set of tiles.
    We use Greedy-Util as the primary baseline because of its systematic approach and similarity to our target use case.
    \item \textbf{3D-aCortex}~\cite{bavandpour2019}: The 3D-aCortex benchmark greedily maps \acp{mvm} with maximized \ac{tp}, repeating the mapping process $n$ times (1000 was chosen here) with randomized \ac{mvm} orderings.
    The mapping with the fewest vertical tiers is then selected.
    \item \textbf{Full-compute}: We propose an \textit{idealized} performance upper bound that represents the potential throughput if the \ac{3dimc} macro's compute resources remain fully utilized throughout inference.
    This is computed by configuring all \acp{mvm} with the maximum degree of \ac{tp} and allowing weight remapping, e.g., allowing any \ac{mvm} to run on any open set of \acp{lg} with no additional cost.
    While such behavior cannot be expected from any mapping algorithm in reality due to the poor write and endurance characteristics of \ac{nvm} devices~\cite{jain2023}, this assumption nevertheless serves as a useful reference, representing the compute roofline of the \ac{3dimc} accelerator.
    
\end{enumerate}

\section{Experimental Results}

In this section, we analyze \ac{camp}'s throughput and robustness across various settings, including the target \ac{llm}, runtime context, system configuration, additional constraints, and end-to-end performance implications.
Supplementary results exploring expanded analyses and ablations are deferred to the Appendix.

\subsection{3D-IMC accelerator performance improvement}

We first analyze \ac{camp}'s throughput benefit within the scope it directly affects: the execution of \acp{mvm} on the \ac{3dimc} accelerator during the decode phase.
Figure~\ref{fig:imc_speedup} presents the throughput results normalized to the Greedy-Util baseline.
% We compare the mapping algorithms across different batch sizes and target \acp{llm}.
% In addition to the mapping algorithms, the figure includes a gray bar labeled \textbf{full compute utilization}, which represents potential throughput if the \ac{imc} macro's compute resources remain fully utilized throughout inference.
% To compute this bound, we construct an idealized mapping that permits unrestricted \ac{tp}, \ac{bp}, and \ac{ep}, even when the mapping is physically infeasible.
% We then serialize computation that exceeds the physical resource limit along the time dimension; for example, if the idealized schedule uses 4{,}000 tiles but the \ac{imc} macro can activate only one tier in each of its 1{,}024 tiles, the excess computation is delayed accordingly.
% The full-compute-utilization setting therefore provides an idealized upper bound on speedup.
% As batch size increases, more opportunities for parallel execution increase potential for improvement.
% The results demonstrate that, when the \ac{imc} macro is fully saturated at sufficiently large batch sizes, optimizing the weight mapping can improve throughput by up to $2\times$.

% At smaller batch sizes, however, the potential for improvement is inherently more limited due to fewer opportunities for parallel execution.
% Nevertheless, \ac{camp} remains much closer to the infinite-tiles bound than the other baselines.

At smaller batch sizes, \ac{bp} is limited, with no \ac{bp} at a batch size of 1.
Accordingly, for dense \acp{llm} such as Llama and Qwen, the \ac{tp}-maximizing baselines are optimal as only \ac{tp} is available, with \ac{camp} matching their performance.
However, for \ac{moe}-based \acp{llm}, \ac{ep} provides an additional source of parallelism even in single-batch workloads.
Consequently, the full-compute bound projects a modest $1.15\times$ speedup for Mixtral inference, while \ac{camp} achieves an average throughput speedup of $1.10\times$, reaching up to $1.14\times$ depending on the experts activated by the batch.
% \ac{camp} does not fully reach the infinite-tiles bound, which is expected because accommodating all expert \acp{mvm} across parallelizable leaf groups requires selecting lower \ac{tp} degrees, thereby increasing the latency of individual expert \acp{mvm}.
% \ac{camp} does not fully reach the Full-compute bound because accommodating all expert \acp{mvm} across parallelizable \acp{lg} at maximum \ac{tp} requires 288 \acp{lg} ($28 \times 8$ for \ac{g+u} and $8 \times 8$ for down-projection experts), whereas the target system provides 64 \acp{lg}.

As batch size increases, improving \ac{bp} becomes favorable.
Throughput gains reach $1.95\times$ for Llama 8B, $1.91\times$ for Llama 70B, and $2.0\times$ for Qwen 72B.
For Mixtral, \ac{camp} achieves a $1.71\times$ average speedup and up to a $1.84\times$ maximum speedup.
% This improvement is primarily enabled by batch parallelism across tokens in a batch.
We observe that \ac{camp} does not always maximize the \ac{tp} degree of each \ac{mvm}, allowing mutually exclusive sets of tiles and \acp{lg} to be allocated to different \acp{mvm}.
Although this increases the compute latency of individual \acp{mvm}, the additional latency is offset by an increase in \ac{bp}.

The gap between \ac{camp} and Full-compute widens as the batch size increases.
The underlying assumption that any \ac{mvm} may occupy any free \ac{lg} becomes more prominent as more \acp{mvm} compete for compute space, resulting in less feasible mappings and loosening the upper bound.
Even so, \ac{camp} eliminates a majority of the throughput gap between the baselines and this theoretical upper bound.

% The gap between the full-compute-utilization bound and our mapping methodology widens at high batch sizes.
% In the actual system, \ac{camp} fully utilizes the available compute resources of the \ac{imc} macro, preventing additional batches from being accommodated concurrently.
% In contrast, the full-compute-utilization setting has no such resource-saturation point and can continue to exploit additional batch-level parallelism.
% The bound therefore becomes increasingly loose at high batch sizes and should not be interpreted as tight in this regime.

\begin{figure}[t]
    \centering
    \includegraphics[width=1\columnwidth]{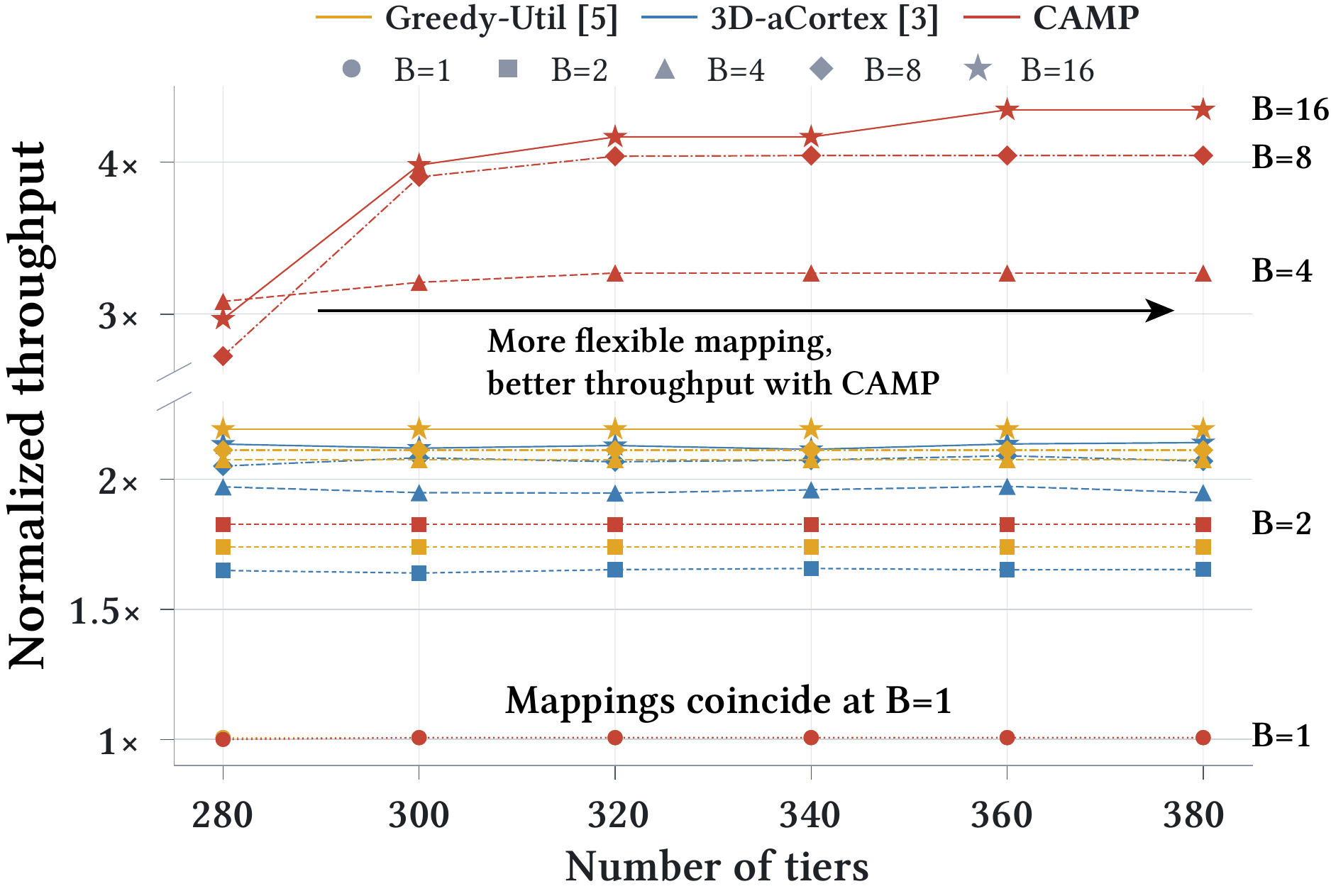}
    \Description{Line chart of normalized throughput versus 280--380 vertical tiers for Greedy-Util, 3D-aCortex, and \acs*{camp} at batch sizes 1, 2, 4, 8, and 16.
    All methods coincide at 1.0 for batch size 1, and both baselines remain nearly flat at every batch size. \acs*{camp} is also flat at batch size 2 but improves as more tiers permit more flexible mappings at larger batches: it rises from about 3.1 to 3.3 at batch size 4, from 2.7 to 4.1 at batch size 8, and from 3.0 to 4.4 at batch size 16.}
    \caption{Normalized Llama 70B throughput vs. vertical tier count across mapping algorithms.}
    \label{fig:tier_limit}
\end{figure}

\subsection{Impact of tier count on throughput}

Although we adopt a 320-tier design~\cite{cho2025}, \ac{3dimc} macros may provide different memory capacities, affecting throughput by limiting mapping flexibility.
Figure~\ref{fig:tier_limit} illustrates this effect.
Limiting the design to 280 vertical tiers reduces throughput at batch sizes 4, 8, and 16, where high \ac{bp} was previously obtained at the cost of some \ac{tp}.
Conversely, relaxing the tier limit to 360 or 380, consistent with ongoing trends in 3D-memory technology, provides additional mapping flexibility and higher throughput.

Greedy-Util and 3D-aCortex, on the other hand, are agnostic to these resource limits.
3D-aCortex always optimizes for the number of occupied tiers and settles on the mapping with the smallest tier occupancy among the sampled candidates, while Greedy-Util simply fills the tiers from bottom to top in \ac{mvm} computation order.
Both therefore reach the same level of throughput regardless of the number of available tiers, with small fluctuations stemming from randomness in the 3D-aCortex algorithm.

\subsection{Robustness across changing runtime context}

\begin{figure}[t]
    \centering
    \includegraphics[width = 0.9\columnwidth]{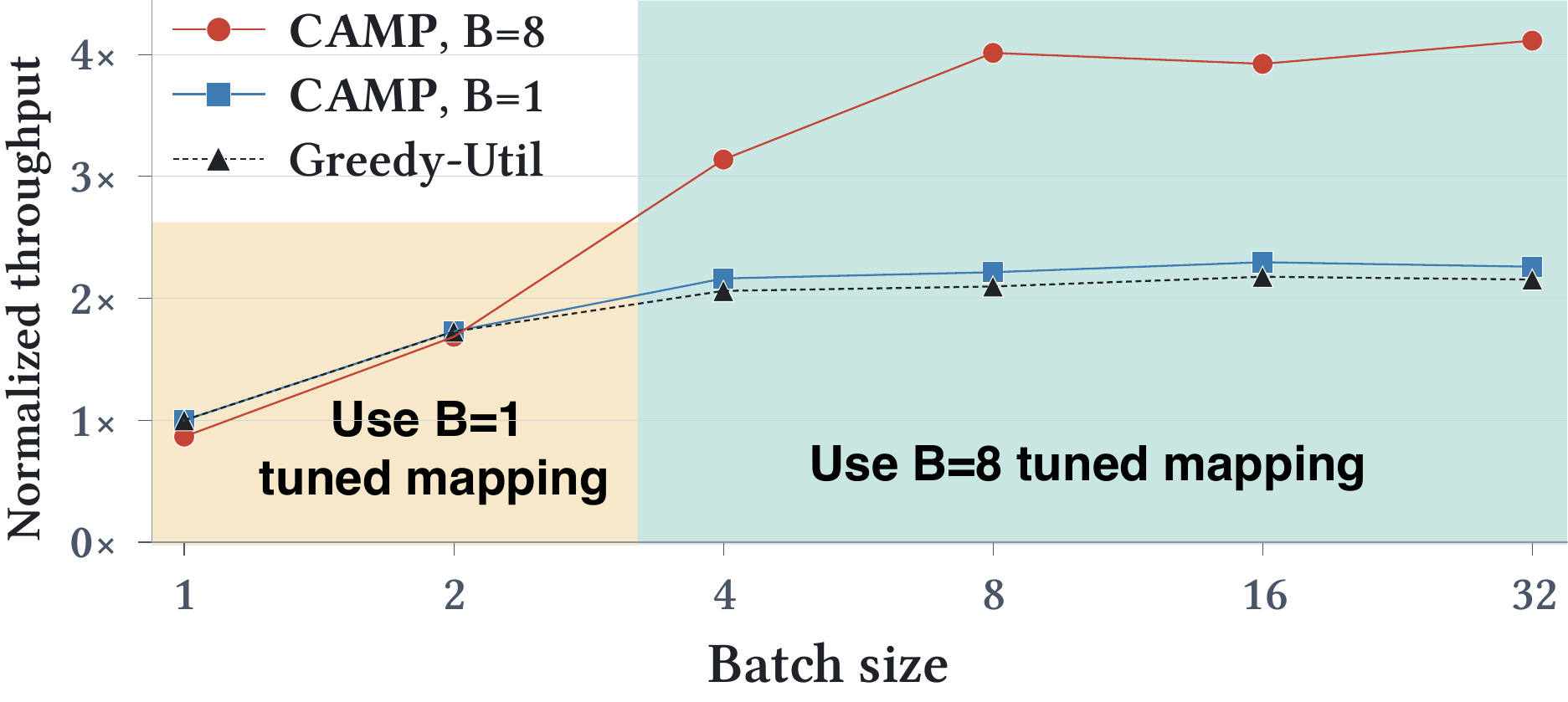}
    \Description{Line chart of normalized throughput versus batch size from 1 to 32.
    The \acs*{camp} mapping tuned for batch size 1 nearly overlaps Greedy-Util, rising from roughly 1.1 at batch size 1 to about 2.5 at batch sizes 8--32.
    The \acs*{camp} mapping tuned for batch size 8 begins at 1.0, matches the other mappings near batch size 2, then rises sharply to 3.6 at batch size 4 and approximately 4.5--4.8 at batch sizes 8--32.
    Shaded regions recommend the batch-1 mapping below batch size 4 and the batch-8 mapping from batch size 4 onward.}
    \caption{Llama 70B speedup with mappings tuned by \acs*{camp} for batch sizes 1 and 8, compared with Greedy-Util. The two mappings remain robust across batch sizes, avoiding remapping and weight programming for new batch sizes.}
    \label{fig:robustness}
\end{figure}

%One of \ac{camp}'s main strengths is its adaptivity to different inference contexts.
Due to the adaptivity of \ac{camp} to different inference contexts, a mapping obtained under the expectation of a single-batch inference differs substantially from one obtained for batch size 8.
This can raise concerns when the inference context is highly variable, i.e., when the batch size changes during \ac{llm} serving.
We therefore analyze robustness in Figure~\ref{fig:robustness}, which shows that a mapping tuned for batch size 8 outperforms Greedy-Util at all batch sizes $\geq 4$.
Thus, with only two mapping options, one tuned for single-batch and one tuned for large batches, all batch sizes can be served more efficiently than with the Greedy-Util alternative.

% Throughput also saturates beyond batch size 8 in the modeled \ac{imc} architecture because the \ac{imc} tiles are nearly fully utilized.
% This also removes the need to tune a separate mapping for larger batch sizes.

% <KPM distances??? Maybe add if we have space>

\subsection{Impact of \texorpdfstring{$W$}{W} and \texorpdfstring{$k$}{k} on CAMP performance}

\begin{figure}[t]
    \centering
    \includegraphics[width=0.9\columnwidth,trim={0cm 0cm 0cm 0cm},clip]{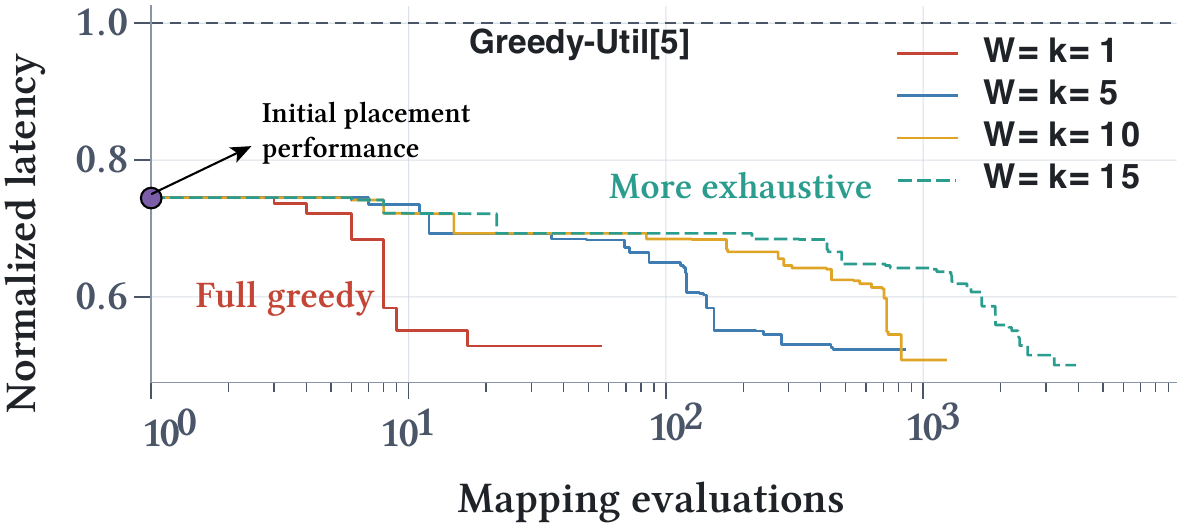}
    \Description{Step curves of normalized latency versus mapping evaluations on a logarithmic horizontal axis for beam-search settings W equals k equals 1, 5, 10, and 15.
    All searches begin at the initial-placement latency of about 0.74 relative to the Greedy-Util reference at 1.0.
    The fully greedy search reaches about 0.52 in fewer than 50 evaluations, while wider searches continue longer and ultimately reach approximately 0.49--0.50; the W equals k equals 15 curve requires several thousand evaluations.}
    \caption{Optimization curves illustrating latency improvement during the refinement phase with different beam search parameters $W$ and $k$.
    The target workload is Llama 70B with a batch size of 8.}
    \label{fig:beam_search_evals}
\end{figure}

\begin{figure*}[htbp]
    \centering
    \includegraphics[width = 2\columnwidth]{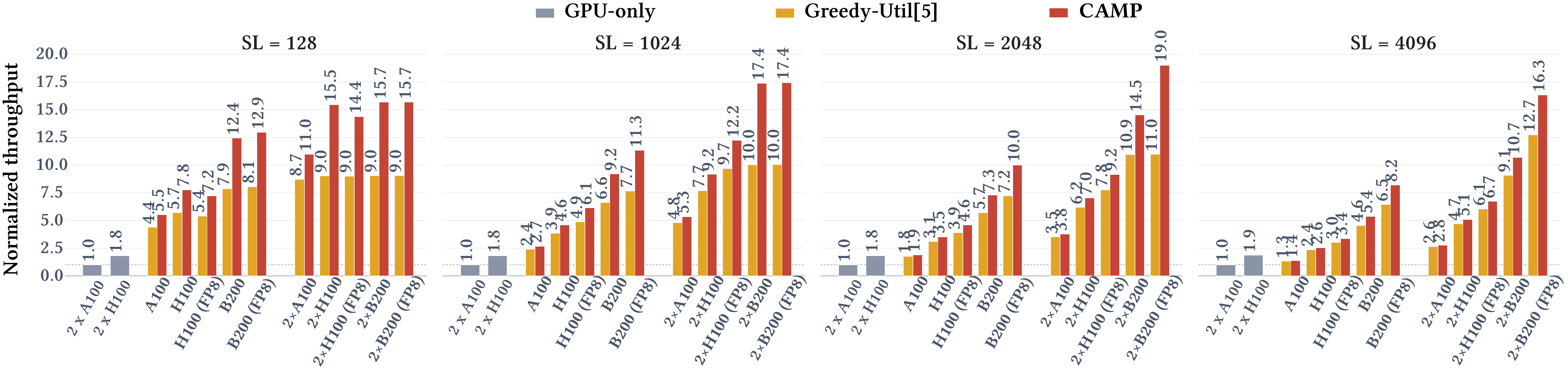}
    \Description{Four grouped-bar panels show end-to-end throughput normalized to two A100 \acsp*{gpu} for sequence lengths 128, 1024, 2048, and 4096.
    Each panel compares \acs*{gpu}-only A100 and H100 baselines with \acs*{3dimc} systems using A100, H100, H100 with FP8 key-value cache, B200, or B200 with FP8, including configurations shared by two \acsp*{gpu}.
    \acs*{camp} exceeds Greedy-Util in every \acs*{3dimc} configuration, with larger absolute gains for faster attention units and shared systems.
    \acs*{camp}'s best speedups are about 15.7, 17.4, 19.0, and 16.3 at the four sequence lengths, versus Greedy-Util maxima of 9.0, 10.0, 11.0, and 12.7.}
    \caption{
        End-to-end decode phase speedup of Llama 70B with a \acs*{3dimc} accelerator paired with various attention units, relative to the 2$\times$A100/H100 \acs*{gpu} baselines.
        Batch sizes of 16/16/64 are utilized for A100/H100/B200 \acsp*{gpu}, respectively.
        In the 2-\acs*{gpu} systems, one \acs*{3dimc} accelerator is shared between two \acsp*{gpu}, each serving the above-mentioned batch sizes.
        }
    \label{fig:end-to-end}
\end{figure*}

% To elaborate on how the beam search converges toward better mapping solutions during the refinement phase, we provide Figure~\ref{fig:beam_search_evals}.
% As discussed in the previous section, for a batch size of 1, \ac{camp} matches the performance of the \ac{tp}-maximizing baseline Greedy-Util~\cite{Buchel2025}, without any additional speedup.
% However, this result is reached through refinement, not at the initial placement.
% The initial placement is fixed to 1-way \ac{tp}, which is a suboptimal choice for single-batch instances.
% Refinement closes the gap by exploring higher \ac{tp} options and committing to them.
% Increasing the beam width $W$ and the action width $k$ does not improve the results here, because the search is straightforward and does not require exhaustiveness.

Figure~\ref{fig:beam_search_evals} illustrates the impact of parameters $W$ and $k$ on \ac{camp} performance when mapping Llama 70B with a batch size of 8.
The first point to note is that maximizing \ac{bp} as performed in the initial mapping phase already provides superior latency over Greedy-Util.
Refinement further improves gains, as the optimal balance between \ac{tp} and \ac{bp} lies somewhere between the two extremes.
Optimization curves indicate that more exhaustive searches enabled by larger $W$ and $k$ values achieve further improved latencies.
We note that these improvements are small at \textasciitilde5\%.
We further note that we terminated the beam search after 400 non-improving action evaluations in these tests, a limit that was chosen empirically and conservatively.
Appendix~\ref{app:beam_search} provides further analysis of beam parameters.

\subsection{Mapping implications on end-to-end latency}
\label{subsec:endtoend}

%Although we show that \acp{mvm} in the decode phase can be substantially accelerated by optimizing weight mapping on the \ac{3dimc} macro, the end-to-end implications of this improvement require further examination.
While the preceding results demonstrate substantial throughput gains for \acp{mvm}, we now examine how these gains propagate to end-to-end decoding performance.
In this section, we present system-level simulation results that pair the \ac{3dimc} accelerator with various \acp{gpu} operating as attention units.
Due to Amdahl's law, a substantial speedup in the \ac{3dimc} accelerator has limited impact on end-to-end performance when \ac{kv} loading and attention computation dominate the inference runtime.
Figure~\ref{fig:end-to-end} presents the end-to-end speedups over 2$\times$A100/H100 \ac{gpu} baselines across \ac{3dimc}-enabled systems complemented with different attention units and sequence lengths for Greedy-Util and \ac{camp}.
With an A100 attention unit and a sequence length of $4096$, weight-mapping optimization provides minimal throughput improvement because attention dominates inference runtime.
Reducing sequence length, for example through \ac{kv}-cache compaction or by targeting shorter-context workloads, increases the benefit: at a sequence length of $128$, \ac{camp} raises the \ac{3dimc} + A100 speedup over baseline to $5.5\times$, compared to $4.4\times$ with Greedy-Util.
More powerful \acp{gpu}, reduced-precision \ac{kv} caching, and sharing the \ac{3dimc} accelerator across multiple \acp{gpu} further balance attention and \ac{mvm} overheads.
When batches from different \acp{gpu} are interleaved for execution of their \acp{mvm} on the \ac{3dimc} accelerator, \ac{aimc}-side utilization increases, making weight-mapping optimization even more important.
Applying these strategies together translates \ac{3dimc} accelerator throughput gains to a $16.3\times$ end-to-end throughput improvement over the \ac{gpu} baseline at a sequence length of 4096, up from $12.7\times$ for Greedy-Util.
Because \ac{kv} loading bottlenecks nearly all inference stacks at long context, many works address it directly through \ac{kv} compression~\cite{zandieh2026,xiao2024,sun2025}, memory-centric accelerators~\cite{gu2025,park2024,heo2024}, and architectural alternatives such as \acp{ssm}~\cite{gu2022}.
As these techniques relieve the \ac{kv} bottleneck, the share of end-to-end latency governed by weight mapping grows, further reinforcing its importance.
Moreover, short-sequence use cases should not be underestimated.
Even today, the \ac{mvm}-bound regimes which benefit from optimized mapping cover important deployments, including edge assistants, short-form \ac{qa}, and classification, where requests are short (e.g., 99\% of Alpaca~\cite{taori2023} prompts contain fewer than 114 tokens).

\section{Related work}

\ac{3dimc} is an emerging computing paradigm whose architectural and system-level design principles are still under active exploration~\cite{Buchel2025,bavandpour2019,shim2021,shim2021_2,dhingra2026,Meng2025,jang2025}.
Despite bringing significant improvements to the \ac{3dimc} domain from architectural~\cite{shim2021}, dataflow~\cite{jang2025}, and application~\cite{dhingra2026} perspectives, a large portion of the prior art does not reveal specific weight mapping details.

Among concrete weight mapping contributions in the literature, 3D-aCortex~\cite{bavandpour2019} stands out as one of the pioneering works in this domain, leveraging 3D-NAND flash arrays for \ac{dnn} acceleration.
In addition, our primary baseline, Greedy-Util, is based on the mapping methodology proposed in~\cite{Buchel2025}.
That work pioneers the use of \ac{3dimc} for \ac{moe} workloads by exploiting the alignment between the sparse computational structure of \acp{moe} and the compute-storage capacity imbalance inherent in \ac{3dimc}.

PackMyWeights~\cite{houshmand2024} is a mapping work in the IMC domain that compares ``stacked'' and ``flattened'' mapping layouts to fit as many weights as possible into SRAM IMC and thereby avoid weight-reloading costs.
HD-MoE~\cite{huang2025} targets 3D-\ac{dram} near-memory processing and optimizes the trade-off between \ac{tp} and \ac{ep} using a \ac{milp}-based method.
As these works are proposed for different memory technologies, their compute constraints and communication characteristics differ from ours, which limits their applicability to 3D-\ac{nvm}-based \ac{aimc}.
In the domain of \ac{nvm}-based \ac{3dimc}, \ac{camp} is, to the best of our knowledge, the first work that targets weight mapping optimization at this scale of robustness and efficiency.

% We adopt its clearly described heuristic-based randomized weight-packing algorithm as one of our baselines.
% Shim and Yu~\cite{shim2021} propose a hierarchical system architecture that leverages 3D-NAND-based subarrays to accelerate \acp{cnn}
% The proposed architecture serves as the foundation for several subsequent studies~\cite{Meng2025,jang2025}.
% NandPIM~\cite{Meng2025} builds upon the hierarchical architecture proposed in~\cite{shim2021} and introduces a genetic-algorithm-based layer-wise quantization scheme that achieves significant performance improvements.
% Jang et al.~\cite{jang2025}, on the other hand, proposes and implements an H-tree based dataflow for single-batch token generation in \acp{llm} due to its superior broadcast and gather capabilities for the large \acp{mvm} encountered in \acp{llm}.

% It proposes a weight-mapping methodology based on a greedy, in-order, utilization-oriented approach.
% Thame~\cite{dhingra2026} is another recent work that exploits the alignment between \acp{moe} and \ac{3dimc}.
% It proposes an efficient architecture that integrates a \ac{tpu} on-chip with \ac{3dimc}, offloading \ac{ffn} experts to the \ac{imc} through attention/\ac{mvm} disaggregation.

% Among these, only Greedy-Util~\cite{Buchel2025} and 3D-aCortex~\cite{bavandpour2019} describe their weight-mapping schemes in sufficient detail to enable reproducibility.
% In this work, we show that their fixed \ac{tp}-maximizing strategies limit robustness under high-batch regimes and \ac{moe}-based \ac{llm} workloads.

\section{Conclusion}

\ac{camp} adapts to the expected \ac{llm} inference context to maximize compute utilization in \ac{3dimc}-based \ac{llm} accelerators by exploiting available tensor-, batch-, and expert-parallelism opportunities.
After encoding the inference context in a conflict matrix, \ac{camp} uses a profiling-driven feedback process to tune weight placement and improve throughput over context-agnostic mapping baselines.
Our experiments show that static \ac{mvm} throughput increases by up to $2.0\times$ and that much of this improvement propagates to end-to-end performance when the system includes efficient disaggregated attention units.
These gains require only an offline software solution and no hardware modifications.

% \section*{Acknowledgments}
% Generative AI was utilized to assist in development of code, generating the visualizations, and for grammatical and typographical proofreading, not for writing.
% The authors performed due diligence in verifying generated output.

% future work todos:
% more top-$k$ expert study; I only analyzed pairwise activation distributions
% higher-dimensional conflict matrix with more representational power
% other search strategies to speed up the refinement search
% optimizations for finding the most robust mapping
% mapping to multi-macro, multi-attention-device systems
% weight replication

\begin{acks}
Generative AI was utilized to assist in development of code and for grammatical and typographical proofreading, not for writing.
The authors performed due diligence in verifying generated output.
\end{acks}

\bibliographystyle{ACM-Reference-Format}
\bibliography{references}


\clearpage

\begin{figure*}[htbp]
    \centering
    \includegraphics[width=0.95\textwidth]{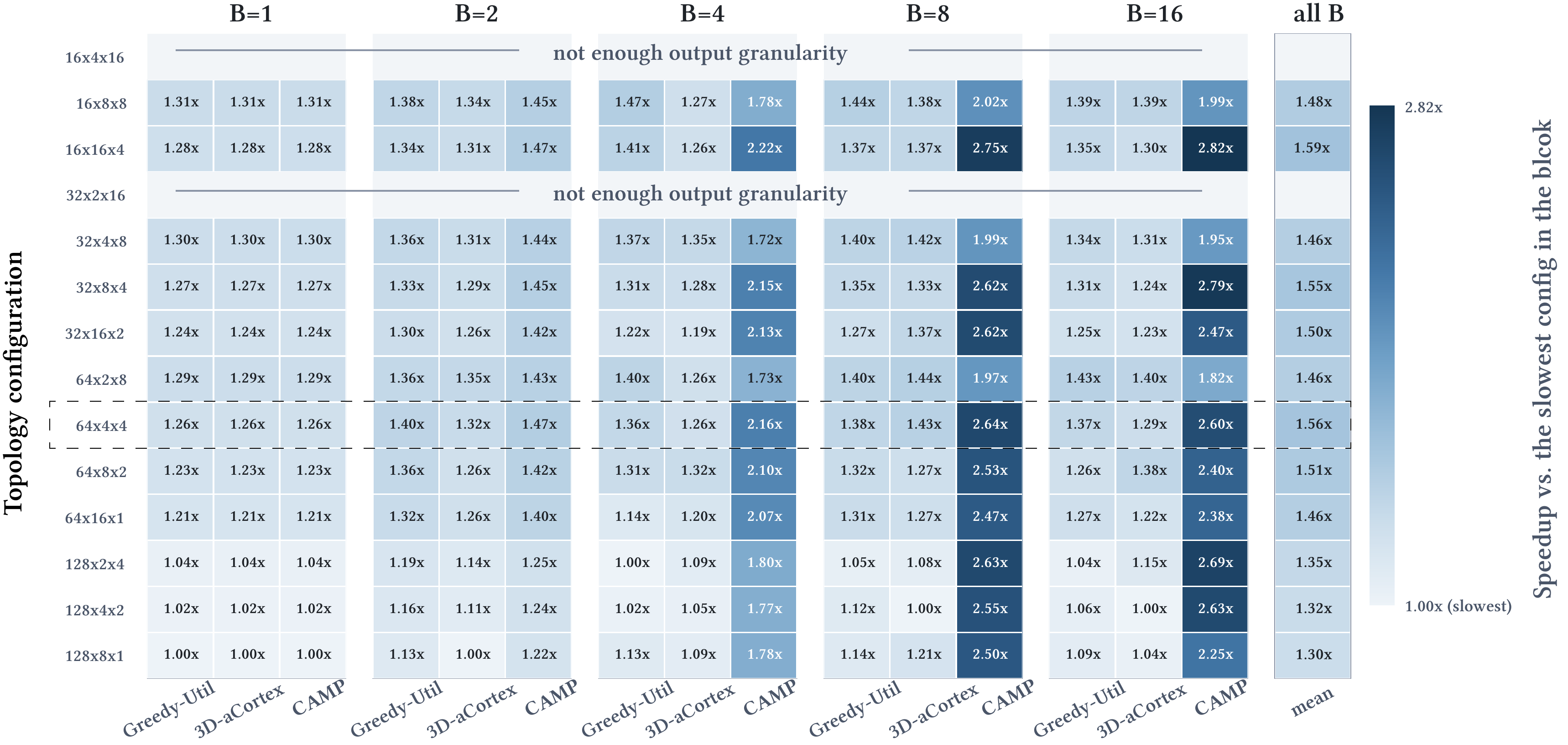}
    \Description{Grid of heatmaps comparing Greedy-Util, 3D-aCortex, and \acs*{camp} across eleven \acs*{3dimc} macro topologies and batch sizes 1, 2, 4, 8, and 16.
    Cell labels give speedup relative to the slowest configuration.
    Two topologies with 16 leaves per tile group are invalid because they lack output granularity.
    Differences are small at batch size 1, while \acs*{camp} is consistently darkest and fastest at larger batches, reaching up to 2.82 times at batch size 16.
    The highlighted 64-by-4-by-4 topology has a mean speedup of 1.56 across batches, near the best overall mean of 1.59.}
    \caption{Throughput comparison for Llama 70B in a grid of \acs*{3dimc} macro topology configurations and mapping algorithms.
     Topology configurations use the format <$N_{LG} \times N_{L} \times N_{T}$>: number of \acsp*{lg}, leaves per \acs*{lg}, and tiles per leaf.}
    \label{fig:imc_topology}
\end{figure*}

\appendix

\section{Architectural characteristics impact analysis}
\label{app:arch_topology}

\begin{figure}[htbp]
    \centering
    \includegraphics[width=0.95\columnwidth]{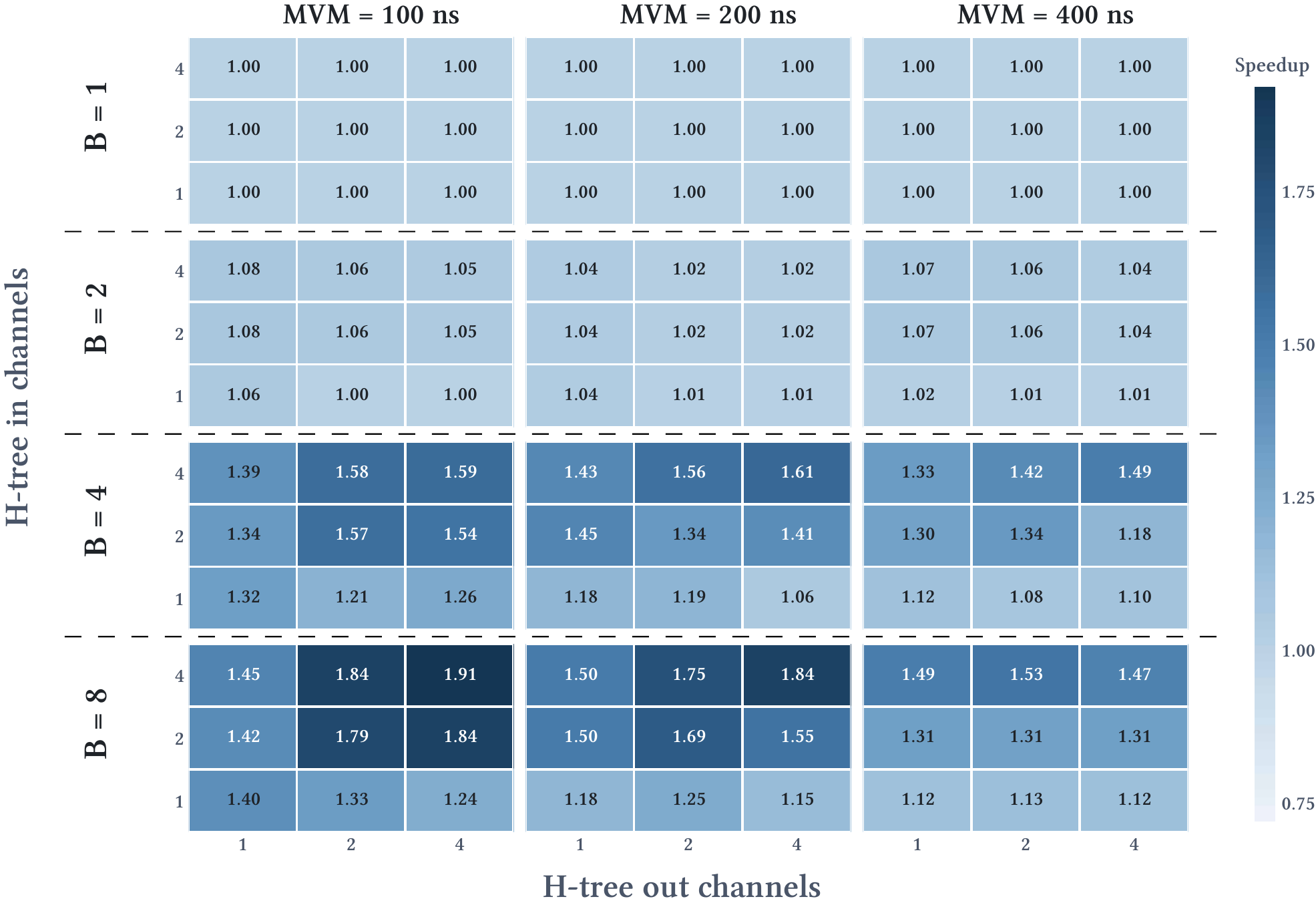}
    \Description{A three-by-four grid of heatmaps reports \acs*{camp} speedup over Greedy-Util.
    Columns group \acs*{mvm} latencies of 100, 200, and 400 nanoseconds and one, two, or four H-tree output channels; row groups use batch sizes 1, 2, 4, and 8, each with one, two, or four H-tree input channels.
    Speedup is exactly one at batch size 1 and remains near one at batch size 2.
    At batches 4 and 8, additional input and output channels generally increase speedup, with a maximum of 1.91 for batch size 8, 100-nanosecond \acsp*{mvm}, four input channels, and four output channels; gains generally shrink for slower \acsp*{mvm}.}
    \caption{Throughput speedup of \acs*{camp} over Greedy-Util~\cite{Buchel2025} as the number of input/output channels and \acs*{mvm} latency vary, profiled on Llama 70B.}
    \label{fig:channel_latency}
\end{figure}

\begin{figure}[htbp]
    \centering
    \includegraphics[width = 0.98\columnwidth]{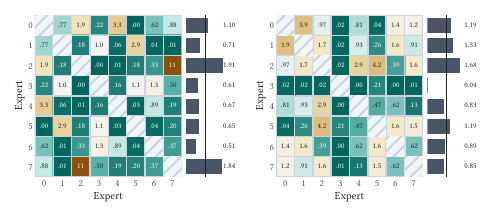}
    \Description{Two 8-by-8 expert co-activation matrices for Mixtral 8-by-7B, with accompanying marginal-frequency bars.
    The left matrix represents global-facts prompts at layer 31 and the right represents business-ethics prompts at layer 11.
    Diagonal entries are zero, while off-diagonal values vary substantially rather than forming a uniform pattern; the two domains emphasize different expert pairs and have different per-expert marginal frequencies.}
    \caption{
        Expert co-activation frequencies with respect to uniform frequency in Mixtral-8$\times$7B for 50 MMLU prompts~\cite{hendrycks2021} from the ``global facts'' (left, layer 31) and ``business ethics'' (right, layer 11) domains.
        Although \acs*{moe}-based networks are trained to balance expert utilization, co-activation frequencies vary across domains and layers.}
    \label{fig:coactivations}
\end{figure}

This appendix analyzes \ac{camp}'s adaptability to varying architectural design choices and constraints.
\subsection{3D-IMC macro topology design space exploration}
\label{app:configsweep}
As indicated in Section~\ref{sec:experimental_setup}, the number of tiles in the \ac{3dimc} macro is fixed at 1{,}024 due to capacity requirements for serving modern \ac{llm} workloads.
However, the hierarchical configuration of the \ac{3dimc} macro, represented as $(N_{LG} \times N_{L} \times N_{T})$, where $N_{LG}$, $N_{L}$, and $N_{T}$ denote the number of \acp{lg}, the number of leaves per \ac{lg}, and the number of tiles per leaf, respectively, remains a flexible design parameter that can have significant performance implications.

The first consideration is \ac{tp} capacity.
Structurally, accumulation is performed within \acp{lg}, where the accumulation units are located.
Increasing the number of tiles within an \ac{lg} allows more partial sums to be computed concurrently, thereby enabling a higher degree of \ac{tp}.
For example, the $(16\times16\times4)$ configuration has 64 tiles per \ac{lg}, whereas the $(128\times4\times2)$ configuration has only 8 tiles per \ac{lg}, resulting in more limited \ac{tp}.

The second consideration is output granularity.
A leaf cannot generate more than one output tensor because the partial sums from the tiles within a leaf are always accumulated.
Therefore, at most $N_{LG} \times N_{L}$ output tensors can be gathered in a single gather step.
Consequently, configurations such as $(16\times4\times16)$ and $(32\times2\times16)$ cannot accommodate models with large \acp{mvm}, such as Llama 70B, whose \acp{mvm} can generate more than 64 output chunks.
Detailed ablation results on these trade-offs are presented in Figure~\ref{fig:imc_topology}.

Based on these considerations, we select the $(64\times4\times4)$ configuration, as it provides competitive performance across a range of workloads while offering a favorable trade-off between \ac{tp} capacity and output granularity.

% \begin{figure*}[]
%     \centering

%     \begin{subfigure}[b]{0.48\textwidth}
%         \centering
%         \includegraphics[width=\textwidth]{figures/topology_mvm_latency_mvm100.pdf}
%         \caption{MVM latency = 100 ns}
%         \label{fig:mvm_latency_ablation_1}
%     \end{subfigure}
%     \hfill
%     \begin{subfigure}[b]{0.48\textwidth}
%         \centering
%         \includegraphics[width=\textwidth]{figures/topology_mvm_latency_mvm200.pdf}
%         \caption{MVM latency = 200 ns}
%         \label{fig:mvm_latency_ablation_2}
%     \end{subfigure}

%     \caption{Given different architecture topology configurations with different numbers of leaf groups, leaves, and tiles on the hierarchy, the change in the MVM latency.}
%     \label{fig:mvm_latency_ablation}
% \end{figure*}

% \section{Impact of TP level on individual MVM compute latencies}

% \begin{figure}[]
%     \centering
%     \includegraphics[width=0.7\columnwidth]{figures/tp_speedups_merged.pdf}
%     \caption{Speedups with increasing -way tensor parallelism (TP) for different input vector dimensions}
%     \label{fig:tp_speedup}
% \end{figure}

\subsection{H-tree bandwidth}
\label{app:bandwidth}
Even when weight mapping enables high \ac{bp} and \ac{ep}, communication bottlenecks in the dataflow can prevent \acp{mvm} from executing concurrently.
To assess the sensitivity of throughput to the H-tree communication bandwidth, we evaluated the system with a varying number of H-tree input and output channels.
The results are shown in Figure~\ref{fig:channel_latency}.
When communication is limited to a single input/output channel, the \ac{bp} enabled by the mapping is masked because inputs cannot be broadcast concurrently.
As can be seen, \ac{camp} throughput gains vary between $1.40\times$ and $1.91\times$ as the channel count is swept from 1 to 4.
This indicates that an optimized mapping must be paired with sufficient H-tree bandwidth to fully utilize compute resources.

\subsection{MVM latency on 3D-IMC tiles}
\label{app:mvmlatency}
\Ac{mvm} latency is the time required to compute a $512 \times 512$ \ac{mvm} on an active tier of a \ac{3dimc} tile.
Depending on the adopted technology, this latency may vary.
\ac{camp} has a greater impact with faster \ac{aimc} tiles.
Figure~\ref{fig:channel_latency} shows that peak throughput gains occur at an \ac{mvm} latency of 100\,ns.
As \ac{mvm} latency increases, \ac{tp} becomes more beneficial.
Because Greedy-Util maximizes \ac{tp} by default and \ac{camp} adapts to prioritize \ac{tp}, the two algorithms converge toward similar mappings.

\begin{figure}[htbp]
    \centering
    \includegraphics[width = 0.95\columnwidth]{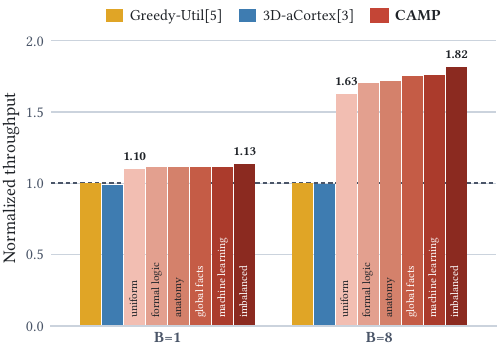}
    \Description{Grouped bars compare normalized throughput for Greedy-Util, 3D-aCortex, and \acs*{camp} at batch sizes 1 and 8.
    Both baselines remain at approximately 1.0 for all expert distributions.
    \acs*{camp} is shown for uniform, formal-logic, anatomy, global-facts, machine-learning, and synthetic imbalanced distributions; its throughput ranges from about 1.10 to 1.13 at batch size 1 and increases monotonically from 1.63 for the uniform distribution to 1.82 for the imbalanced distribution at batch size 8.}
    \caption{Speedup across expert-activation distributions derived empirically from input domains in the MMLU dataset~\cite{hendrycks2021}.
    ``Uniform'' and ``imbalanced'' are synthetic distributions illustrating the extremes. Uniform is fully uniform on expert-co-activations. And for ``imbalanced'', we increased \textit{coefficient of variation} (CV = std/mean) for co-activation probabilities to 0.8 where the realistic domains from MMLU have it between 0.15 and 0.32.
    Results are averaged over 50 trials.}
    \label{fig:domain_specific}
\end{figure}

\section{Domain-specific expert parallelism in MoEs}
\label{sec:expert_imbalance}

Even though the \ac{moe}-based networks are trained for uniform utilization of expert \acp{mvm}, it is possible to encounter imbalanced utilization of experts at the layer level when prompts from specific domains are used.
Figure~\ref{fig:coactivations} illustrates this on two example domains.
Domain-specific expert-activation imbalances are represented in the mapping process via the expanded conflict matrix in Equation~\ref{eq:moe_conflict_expansion}.
The weighting function $\omega$ uses the joint distribution of expert-activation combinations, computed by averaging over domain specific prompts from the MMLU dataset~\cite{hendrycks2021}.

Given the set of experts $\mathcal{E}$ with $|\mathcal{E}| = E$, of which the top $K$ experts are activated per token per layer, let $R\subseteq\mathcal{E}$ with $|R|=K$ denote one of $\binom{E}{K}$ such combinations.
We specify $P_i(R)$ as the probability that subset $R$ is selected for a token at the undecomposed \ac{mvm} $i$.
This distribution can be estimated empirically from a representative set of input prompts: the model is executed over the dataset, the activated expert subset is recorded for each token, and the resulting frequencies are normalized to obtain $P_i(R)$.
Individual and pairwise expert activation probabilities then follow by marginalization:

\begin{equation}
    p_{k} = p_{i,e} = \sum_{R \subseteq \mathcal{E} \wedge e \in R} P_i(R)
\end{equation}

The corresponding pairwise probabilities are:
\begin{equation}
    p_{k,l} = p_{(i,e_1),(i,e_2)} = \sum_{R \subseteq \mathcal{E} \wedge e_1,e_2 \in R} P_i(R)
\end{equation}

for experts $e_1$ and $e_2$ of \ac{mvm} $i$.
Since the conflict matrix represents pairwise conflicts only, no higher-order marginals are required.
Given these probabilities, the weights are:

\begin{equation}
    \omega(k,l) =
    \begin{cases}
        p_{k,l} & k = (i,e) \wedge l = (i,e') \\
        p_k \cdot p_l & \text{otherwise}
    \end{cases}
\end{equation}

with the convention $p_k = 1$ when $k$ is not an expert \ac{mvm}.
The intuition is to use measured co-activation statistics where they matter most, namely between sibling experts of the same \ac{mvm}, and to fall back on an independence assumption for inter-\ac{mvm} conflicts.
This process produces the differences between the conflict matrices in Figure~\ref{fig:conflict_matrices}.

We evaluate \ac{camp}'s robustness across expert-activation distributions induced by different prompt domains.
Figure~\ref{fig:domain_specific} shows that greater expert-activation imbalance produces larger speedups.
The speedup over Greedy-Util rises from $1.63\times$ to $1.82\times$ when we switch to an imbalanced distribution.
This imbalance indicates which expert \acp{mvm} should occupy mutually exclusive resources and which can share resources.
On realistic samples from the MMLU dataset~\cite{hendrycks2021}, speedups range from $1.63\times$ to $1.82\times$, reflecting the limited imbalance present.
Greedy-Util performance does not vary across domains because it does not enable \ac{ep}: when all expert \acp{mvm} are serialized, the activated experts do not affect performance.
3D-aCortex fluctuates slightly, but the variation is within run-to-run noise because it minimizes tier usage without considering expert activations.

\section{Mapping search space and runtime complexity analysis}
This appendix analyzes the bounds of the search space that contains all mapping possibilities and analyzes the runtime complexity of the \ac{camp} algorithm.

\subsection{Bounds of the search space cardinality}
\label{sec:search_space_cardinality}

\begin{figure}[htbp]
    \centering
    \includegraphics[width=0.95\columnwidth]{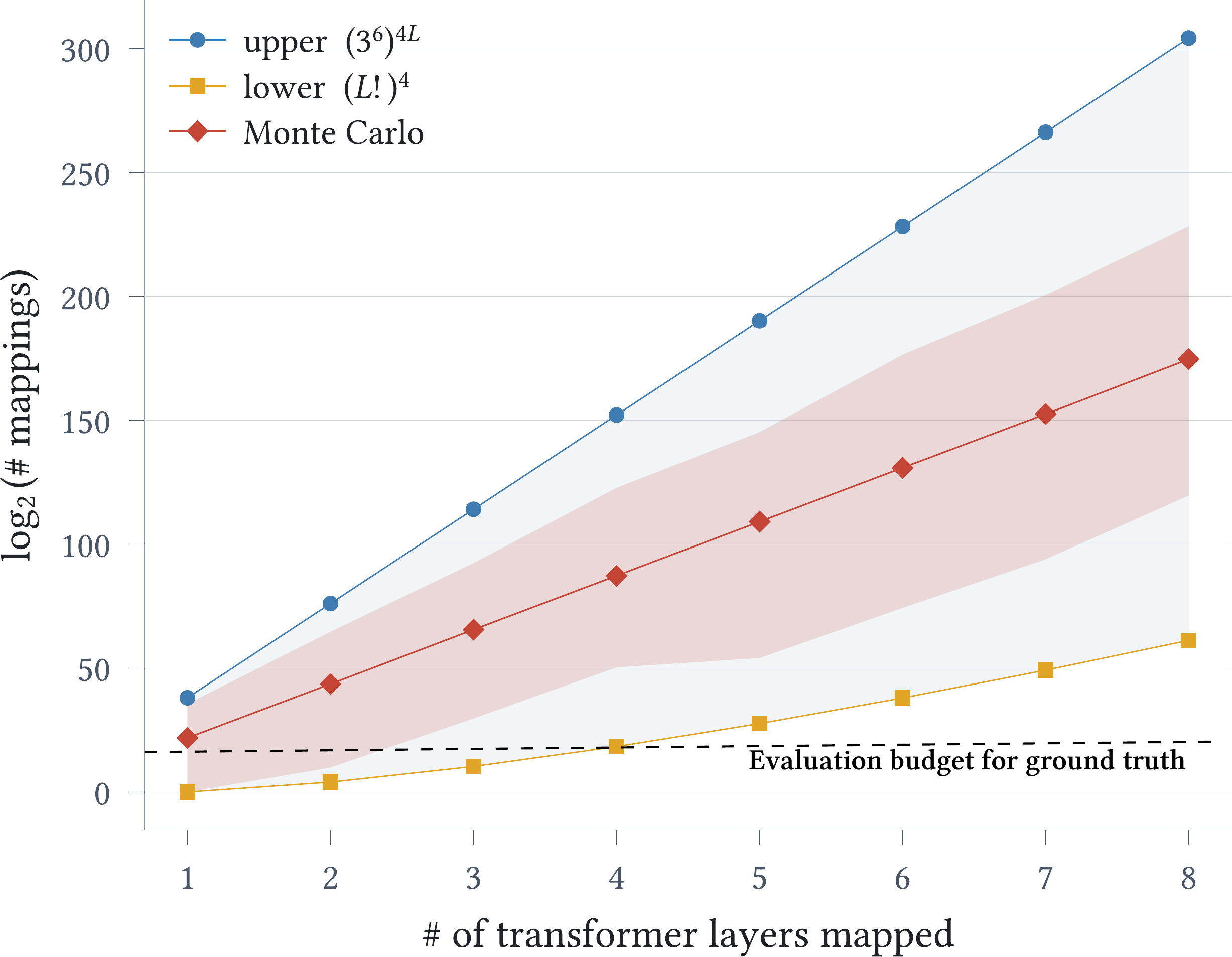}
    \Description{Line chart of the base-2 logarithm of the number of mappings versus one through eight mapped transformer layers.
    The conservative upper bound grows from about 38 to 304, the Monte Carlo estimate grows from about 22 to 175 with a broad uncertainty band, and the lower bound grows from zero to about 61.
    A dashed horizontal line near 18 marks the evaluation budget for obtaining ground truth and is exceeded by the estimate even for one layer.}
    \caption{Estimated number of options for mapping $L$ Llama 70B transformer layers onto the modeled \acs*{3dimc} macro.
    The red curve shows the Monte Carlo estimate; the other curves show conservative lower and upper bounds.}
    \label{fig:search_space_bounds}
\end{figure}

The size of the mapping search space motivates a search-space-pruning algorithm such as \ac{camp}.
The search space can be compared to a 3D bin-packing problem, in which \acp{mvm} represent reshapable 3D boxes that must be packed into a single bin.
While exact enumeration of the mapping space is intractable, we characterize its complexity using conservative lower and upper bounds and an unbiased Monte Carlo estimator.
Let $|S|$ denote the mapping-space cardinality.

\noindent\textbf{Lower bound:}
The lower bound can be intuitively described as selecting any mapping such that all \acp{mvm} fit within the bin, and then permuting the location of all identically sized \acp{mvm}.
The total number of possible permutations represents the lower bound of the search space.

Consider mapping the weights of the set of \acp{mvm} $\mathcal{J}$ from $L$ transformer layers.
For a conservative bound, we fix the \ac{tp} degree of each \ac{mvm} in $\mathcal{J}$ and partition the \acp{mvm} into equivalence groups by size.
Let $g_m$ denote the group of \acp{mvm} of size $m$, and let $G$ denote the set of all groups.
Then:

\begin{equation}
    |S| \geq \prod_{g_m \in G} |g_m|!
    \label{eq:s_lower_bound}
\end{equation}

Given any feasible mapping of all \acp{mvm} in $\mathcal{J}$, the placements of same-sized \acp{mvm} can always be exchanged.
This gives $|g_m|!$ permutations for each group $g_m$.
Combining the permutations across groups yields Equation~\ref{eq:s_lower_bound}.

For Llama 70B, each transformer layer contains four distinct sizes of \acp{mvm}.
Due to the repetitive structure of transformers, these \acp{mvm} repeat for $L$ layers.
When mapping $L$ layers of Llama 70B, there are four equivalence groups with $L$ \acp{mvm} in each.
Equation~\ref{eq:s_lower_bound} becomes:

\begin{equation}
    |S| \geq (L!)^4
\end{equation}

This lower bound only permutes same-sized \acp{mvm} among their placements; it excludes alternative \ac{tp} degrees and placement layouts.
Even this loose lower bound shows that mapping eight layers of Llama 70B requires at least $(8!)^4 \approx 2^{61}$ evaluations to identify the optimum.

\noindent\textbf{Upper bound:}
System capacity is the limiting factor on the mapping space.
To obtain a conservative upper bound on mapping-space cardinality, we assume that the system has no vertical-tier limit.
Intuitively, this removes inter-\ac{mvm} constraints, allowing every \ac{mvm} to assume all possible shapes.
Thus, the permutation of all possible shapes for all possible \acp{mvm} represents the upper bound of the mapping search space.

Each \ac{mvm} in $\mathcal{J}$ can be mapped to any subset of \acp{lg} on the \ac{3dimc} macro, constrained only by H-tree broadcasting: at each level, the input may propagate left, right, or both, thereby determining the single-pass reachable \acp{lg} for weight mapping~\cite{deng2020}.
Accordingly, there can be $3^d$ such subsets, where $d$ is the depth of the H-tree.
Then, if the \acp{lg} in a subset are already occupied, the weights are simply mapped to one tier above.
In this setting, the mapping search space consists of $|\mathcal{J}|$ independent decisions, yielding the upper bound:

\begin{equation}
    |S| \leq {(3^d)}^{|\mathcal{J}|}
\end{equation}

For Llama 70B, $|\mathcal{J}| = 4L$, and for our architectural setting, $d = 6$.

\noindent\textbf{Monte Carlo estimator:}
In reality, the size of the search space changes as \acp{mvm} are mapped, and future \acp{mvm} are constrained.
Thus, the true search space falls somewhere between the lower and upper bounds.

We quantify this via a Monte Carlo simulation that randomly orders the \acp{mvm} in $\mathcal{J}$ and maps them one by one.
The number of mapping options for \ac{mvm} $j \in \mathcal{J}$ is denoted by $n_j$.
A mapping is picked uniformly at random and committed.
The mapped resources are marked as occupied to obey the occupancy constraints.
The number of mapping options for a trial is the product $\prod_{j \in \mathcal{J}} n_j$ of the option counts at each placement step:
\begin{equation}
    X = \prod_{j = 0}^{|\mathcal{J}| - 1} n_j
\end{equation}

$X$ is an unbiased estimator of $|S|$:
\begin{equation}
    \mathbb{E}[X] =
    \sum_{\mathbf{M} \in S} \big( \prod_{j = 0}^{|\mathcal{J}| - 1} \frac{1}{n_j}  \big) \times \big( \prod_{j = 0}^{|\mathcal{J}| - 1} n_j \big) =
    \sum_{\mathbf{M} \in S} 1 = |S|
\end{equation}
Reaching a specific leaf in the Monte Carlo search tree has probability $\prod_{j = 0}^{|\mathcal{J}| - 1} \frac{1}{n_j}$, and the reached leaf returns $\prod_{j = 0}^{|\mathcal{J}| - 1} n_j$ mapping options.
We report the average across trials as the estimated search-space cardinality.

The red Monte Carlo curve in Figure~\ref{fig:search_space_bounds} lies between the lower and upper bounds and confirms that exhaustive search is infeasible.
Specifically, identifying the optimal mapping for eight layers of Llama 70B requires approximately $2^{175}$ simulation evaluations.
The search space grows further for the \ac{moe}-based Mixtral network because decomposing expert \acp{mvm} yields 18 \acp{mvm} per transformer layer rather than four.

Although exact optimization methods such as \ac{milp} may appear applicable, they require proxy objective functions.
Developing proxy objectives that capture batch-level overlap and resource-conflict interactions remains future work.

\begin{figure*}[htbp]
    \centering
    \includegraphics[width=1.95\columnwidth]{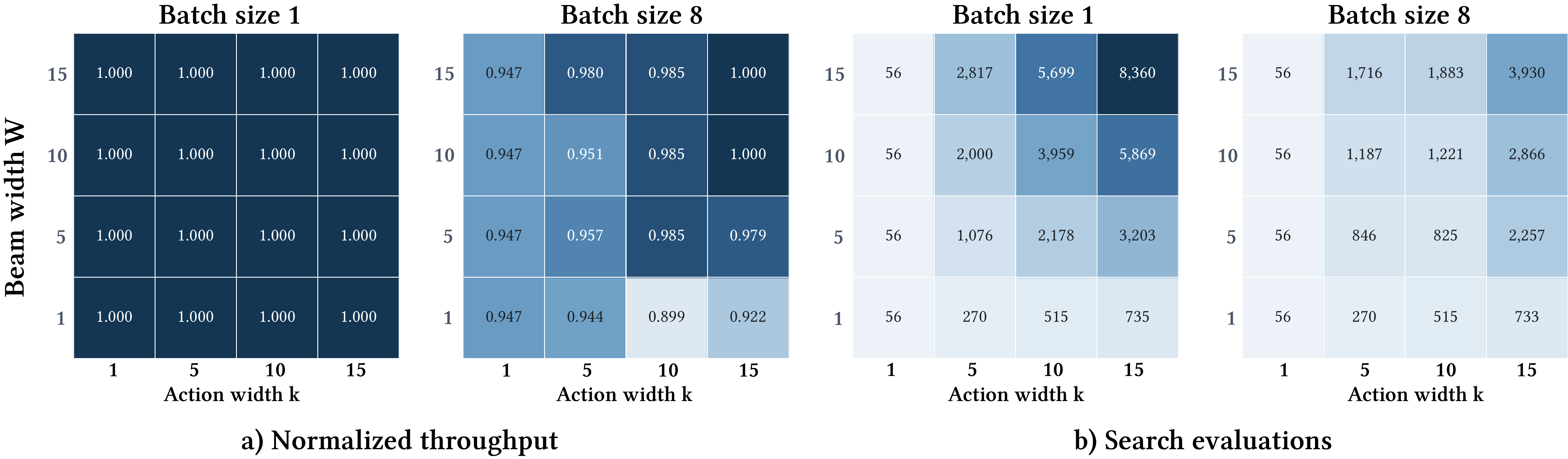}
    \Description{Four four-by-four heatmaps sweep beam width W and action width k over 1, 5, 10, and 15.
    The two throughput heatmaps show that every setting reaches normalized throughput 1.0 at batch size 1, while batch-size-8 throughput ranges from 0.899 to 1.0 and generally improves with wider beams and actions.
    The two search-evaluation heatmaps show rapidly increasing cost: evaluations range from 56 at W equals k equals 1 to 8,360 at batch size 1 and 3,930 at batch size 8 when both widths equal 15.}
    \caption{Ablation on parameters of beam width $W$ and top $k$ actions.}
    \label{fig:wk_ablation}
\end{figure*}

\begin{figure*}[htbp]
    \centering
    \includegraphics[width=\textwidth]{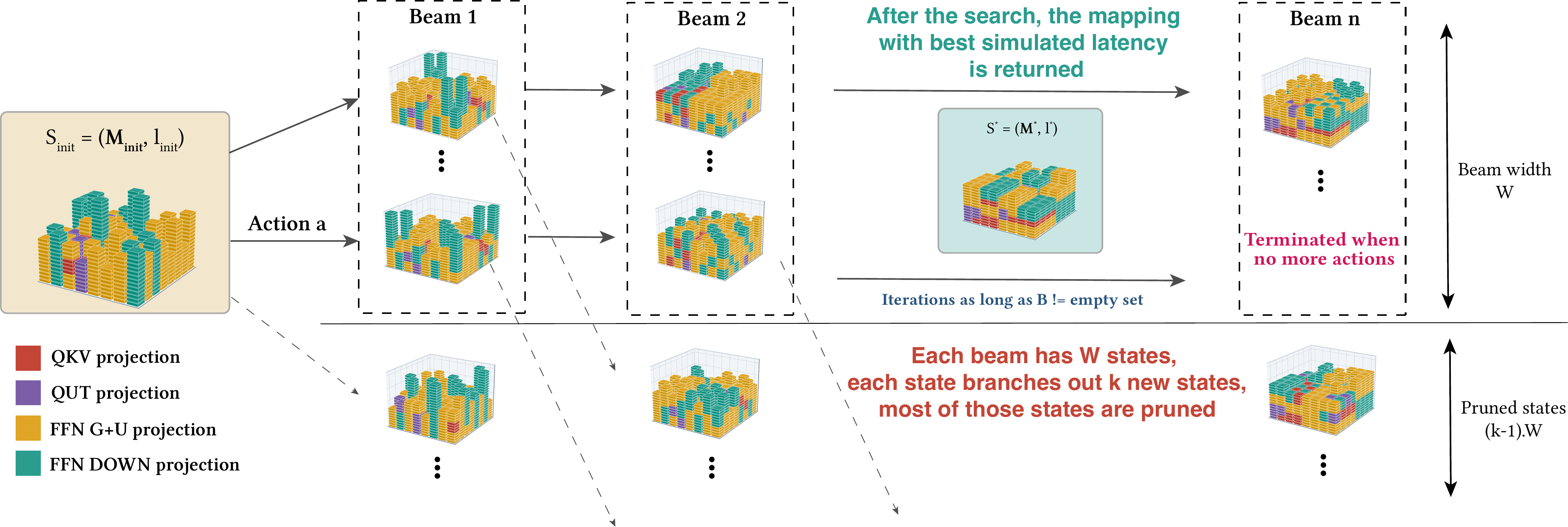}
    \Description{Beam-search schematic.
    An initial state containing mapping M-init and latency l-init branches through refinement actions into W candidate mapping states in Beam 1.
    Each retained state branches into k new states in the next beam; only W states remain above the horizontal divider while the other open-parenthesis k minus 1 close-parenthesis times W states are pruned below it.
    Beams continue until no actions remain, and the mapping with the best simulated latency among all visited states is returned.
    Colored stacks identify \acs*{qkv}, output, \acs*{g+u}, and down-projection placements.}
    \caption{Refinement iterations with beam search.}
    \label{fig:beam_search}
\end{figure*}

\subsection{Impact of \texorpdfstring{$W$}{W} and \texorpdfstring{$k$}{k} on throughput and search space coverage}
\label{app:beam_search}

As explained in Section~\ref{sec:refinement}, we adopt a \textbf{beam search} strategy when selecting a good set of \ac{tp} degrees for the \acp{mvm}, with the beam width $W$ and the action width $k$ serving as tunable parameters that control the exhaustiveness of the search.
In the previous section, we also showed the scale of the search space and the impracticality of a fully exhaustive search.

The question that remains is how exhaustive the beam search must become to close the optimality gap between the baselines and the idealized throughput upper bound for the modeled \ac{3dimc}-based \ac{llm} accelerator.
Figure~\ref{fig:wk_ablation} shows the relative throughput obtained and corresponding number of mapping evaluations as the $W$ and $k$ parameters are ablated.
When the batch size is 1, the optimization path is straightforward, because the most performant mapping is to maximize the degree of \ac{tp} for all mapped \acp{mvm}.
There is therefore no throughput difference between a fully greedy beam search ($W = k = 1$) and more exhaustive configurations.
Batch size 8, on the other hand, requires balancing the degrees of \ac{tp} and \ac{bp}, so increasing $W$ and $k$ generally improves the throughput reached by the beam search.
We report our results with the $W = k = 5$ configuration in the other figures and results; we note, however, that they could be improved by $4\%$ had we chosen the more exhaustive option of $W = k = 15$.
This comes at the cost of roughly $4.6\times$ more mapping evaluations.
Nevertheless, even $W = k = 15$ leads to only 3930 evaluations, a relatively low number with respect to the search-space complexity shown in the previous section.

One exceptional data point is $W = 1, k = 10$, which results in a decrease in throughput compared to the lower action width configurations at the same beam width.
The reason is that increasing the action width at $W = 1$ contributes little to the exhaustiveness of the search.
The most latency-improving action is selected among 10 candidates instead of 1 or 5, but this only makes the search a \textit{steepest descent}, which is still a purely greedy approach.
Getting stuck in a worse local minimum therefore remains possible unless the beam width $W$ is increased.

\subsection{Runtime complexity of CAMP}
\label{app:runtime_complexity}

\ac{camp} consists of three phases with different complexities.
Let $N$ denote the number of \acp{mvm} to be mapped.
The profiling phase maps the \acp{mvm} under the assumption of infinite compute resources, so the mapping is a single greedy pass in which each \ac{mvm} is placed one by one.
The complexity of the profiling phase is therefore $O(N)$.
The initial mapping phase assigns 1-way \ac{tp} to every \ac{mvm} and places them one by one in the same greedy, in-order fashion.
The conflict matrix obtained from the profiling phase is consulted during the placement of each \ac{mvm}, but this adds little cost because the matrix is sparse row-wise.
Denoting the average number of non-zero entries per row as $C$, the complexity of the initial mapping phase is $O(CN)$, which reduces to $O(N)$ when $C$ is independent of $N$.
The refinement phase is the most time-consuming part of the algorithm.
A placement procedure similar to the one in the initial mapping phase is applied iteratively until no further actions remain.
At each refinement step, $W \times k$ candidate actions are evaluated, where $W$ is the beam width and $k$ is the action width, and each candidate requires a placement pass costing $O(N)$; a single step therefore costs $O(WkN)$.
Each \ac{mvm} admits at most $A$ \ac{tp}-changing actions, so the number of steps is at most $AN$.
For instance, with 1-, 2-, and 4-way \ac{tp}, the possible transitions per \ac{mvm} are 1-to-2 and 2-to-4, giving $A = 2$.
Treating $W$, $k$, and $A$ as constants, the refinement phase costs $O(WkAN^2) = O(N^2)$.
The overall complexity is thus $O(N) + O(CN) + O(WkAN^2) = O(N^2)$ with respect to the number of \acp{mvm} being mapped, $N$.

As mentioned in Section~\ref{sec:simulation}, $N$ can be lowered by mapping fewer transformer layers at once and replicating the mapping for the rest of the layers.
This is a tradeoff between mapping quality and runtime.
Another similar tradeoff is tuning the $W,k$ parameters as shown in Section~\ref{app:beam_search}.

Technically $N$ is not a large number.
For example, for Llama 70B with 80 transformer layers $N = 320$.
For Mixtral, when we decompose expert layers into individual \acp{mvm}, $N = 576$.
Quadratic complexity is not an issue at this scale.
However, each iteration in the refinement step requires invoking the simulator for profile-guided feedback, and simulators can have various latencies depending on their granularity and software stack efficiency.
Therefore, we note that the runtime of the weight mapping algorithm is also determined by the runtime of the simulator.

% As mentioned in Section~\ref{sec:simulation}, $N$ can be reduced by mapping fewer transformer layers at once and replicating the mapping for the remaining layers.
% This is a tradeoff between mapping quality and runtime.
% Tuning the $W$ and $k$ parameters, as shown in Section~\ref{app:beam_search}, offers a similar tradeoff.
% In practice, $N$ is not large.
% For Llama 70B with 80 transformer layers, $N = 320$; for Mixtral, decomposing the expert layers into individual \acp{mvm} gives $N = 576$.
% Quadratic complexity is therefore not a concern at this scale.
% However, each refinement iteration invokes the simulator for profile-guided feedback, and simulator latency varies with modeling granularity and implementation efficiency.
% The runtime of the weight mapping algorithm is thus also determined by the runtime of the simulator.

Finally, we note that weight mapping and remapping on \ac{nvm}-based \ac{3dimc} are restricted due to endurance limitations.
In practice, the weights are programmed once, offline, and then used repeatedly during inference.
Weight mapping therefore does not have to meet strict service-level objectives.
With quadratic complexity, \ac{camp} balances mapping quality against runtime, sitting between fully greedy methods and exact optimization methods of exponential complexity.

\section{GPU Profiling}
\label{sec:gpu_profiling}

End-to-end \ac{gpu}-only latencies are measured with vLLM~\cite{kwon2023} (version 0.18.0) serving
an INT8 weight- and activation-quantized variant of
Llama~3.1~70B~Instruct\footnote{\url{https://huggingface.co/neuralmagic/Meta-Llama-3.1-70B-Instruct-quantized.w8a8}}, at a tensor parallelism degree of two.
Chunked prefill is enabled, prefix caching is disabled, and the fraction of device memory reserved for the engine (\texttt{gpu\_\allowbreak memory\_\allowbreak utilization}) is set to $0.9$.
Disabling prefix caching ensures that every request performs the full amount of work implied by its input sequence length.

For each combination of input sequence length and batch size, the maximum model length (\texttt{max\_model\_len}) and the maximum number of batched tokens per scheduler step (\texttt{max\_num\_batched\_tokens}) are computed dynamically so that all prefill requests of the batch are admitted in a single scheduler step.
The prefill phase therefore completes before generation begins, and no scheduler step mixes prefill and decode work.

We enable the \texttt{torch} profiler at engine initialization and parse the resulting traces offline.
From each trace we retain only the decode-only scheduler steps which can be identified by a context length of zero and a number of generation tokens equal to the batch size.
The reported latency is the mean over ten such steps.

\end{document}